\RequirePackage[2020-02-02]{latexrelease}

\documentclass[%
reprint,
amsmath,amssymb,
aps,
prc,
]{revtex4-2}

\makeatletter
\DeclareRobustCommand\ref{%
	\@ifstar\@refstar\T@ref
}%
\DeclareRobustCommand\pageref{%
	\@ifstar\@pagerefstar\T@pageref
}%
\makeatother

\usepackage{graphicx}
\usepackage{dcolumn}
\usepackage{bm}
\usepackage{xcolor}
\usepackage[colorlinks = true,
linkcolor = blue,
urlcolor  = blue,
citecolor = blue,
anchorcolor = black]{hyperref}  
\usepackage{physics}
\usepackage{epstopdf}
\usepackage{epsfig}
\usepackage{longtable}
\usepackage{comment}
\usepackage{dcolumn}

\usepackage{tablefootnote}

\begin{document}

\preprint{APS/123-QED}
	
\title{Self-consistent calculations for atomic electron capture}
	
\author{V. A. Sevestrean$^{1,2,3}$}

\author{O. Ni\c{t}escu$^{1,2,3}$}
\email{ovidiu.nitescu@nipne.ro}

\author{S. Ghinescu$^{1,2,3}$}

\author{S. Stoica$^{1}$}


\affiliation{$^{1}$ International Centre for Advanced Training and Research in Physics,\\ P.O. Box MG12, 077125 M\u{a}gurele, Romania}

\affiliation{$^{2}$Faculty of Physics, University of Bucharest, 405 Atomiştilor, P.O. Box MG-11, RO-077125, Bucharest-M\u{a}gurele, Romania}

\affiliation{$^{3}$“Horia Hulubei” National Institute of Physics and Nuclear Engineering, 30 Reactorului, POB MG-6, RO-077125 Bucharest-M\u{a}gurele, Romania}

\date{\today}
	
	\begin{abstract}

	We present a comprehensive investigation of electron capture (EC) ratios spanning a broad range of atomic numbers. The study employs a self-consistent computational method that incorporates electron screening, electron correlations, overlap and exchange corrections, as well as shake-up and shake-off atomic effects. The electronic wave functions are computed with the Dirac-Hartree-Fock-Slater (DHFS) method, chosen following a systematic comparison of binding energies, atomic relaxation energies and Coulomb amplitudes against other existing methods and experimental data. A novel feature in the calculations is the use of an energy balance employing atomic masses, which avoids approximating the electron total binding energy and allows a more precise determination of the neutrino energy. This leads to a better agreement of our predictions for capture ratios in comparison with the experimental ones, especially for low-energy transitions. We expand the assessment of EC observables uncertainties by incorporating atomic relaxation energy uncertainties, in contrast to previous studies focusing only on Q-value and nuclear level energies. Detailed results are presented for nuclei of practical interest in both nuclear medicine and exotic physics searches involving liquid Xenon detectors ($^{67}\mathrm{Ga}$, $^{111}\mathrm{In}$, $^{123}\mathrm{I}$, $^{125}\mathrm{I}$ and $^{125}\mathrm{Xe}$). Our study can be relevant for astrophysical, nuclear, and medical applications.

	\end{abstract}
	
	\maketitle
	

\section{\label{sec:level1}Introduction}


Electron capture (EC) is a process in which a proton absorbs an electron from atomic shells and transforms into a neutron, emitting a neutrino of well-defined energy. It is a low-energy process that occurs in atoms with neutron-deficient nuclei, by which the nucleus lowers its atomic number by one unit and is accompanied by energetic rearrangement processes. Thus, if the final nucleus remains in an excited state, it de-excites to the ground state either by a $\gamma$ cascade or internal conversion. Also, as long as the captured electron is not from the outermost shell, the final atom remains in an excited state leading to a rearrangement of the electron shells by emission of X-rays or Auger electrons. These rearrangement processes are essential in the measurement of the EC rates, which relies on the detection of the $\gamma$ and X-ray photons and Auger electrons. It becomes difficult in the absence of $\gamma$ photons when the final nucleus remains in the ground state and for light atoms where the X-ray photons and Auger electrons have very low energies. The continuous improvement of the experimental techniques for measuring EC probabilities also stimulates the improvement of their models and computational methods. The theoretical support for understanding the EC processes is much needed, first to explain the experimental data and then to provide data for many EC transitions not yet measured.

EC processes occur in a wide isotope range of naturally-occurring elements, from beryllium to bismuth, and it was also observed in heavier artificial elements~\cite{Gates2008PRC}. Chief among practical applications are radionuclide metrology~\cite{Broda_2007} and nuclear medicine~\cite{buchegger2006auger,bezak2012atomic,PirovanoNMB2021,KuAnthonyEJNMMI2019}. This is because most of the Auger electrons emitted in EC processes have a few keV kinetic energy, thus depositing within a small range. This makes them a high-accuracy and well-controlled tool for internal radiotherapy, irradiating the specific site of the tumor. This application requires precise information on the decay data, such as radiation energy, emission probabilities, decay modes, and half-life, with which the optimal dose can be obtained. Excellent candidates for this purpose are $^{67}\mathrm{Ga}$, $^{111}\mathrm{In}$, $^{123}\mathrm{I}$, $^{125}\mathrm{I}$.

EC also holds a significant position in various fundamental research studies, including neutrino mass scale determination~\cite{HOLMESexperimentJLTP2016,NuMECsexperimentJLTP2016,ECHoTEPJST2017} and nuclear astrophysics~\cite{Langanke_2021}. Detailed description of the EC processes has also become important, in recent years, for a precise background characterization in exotic searches. Specifically, in liquid xenon experiments, the occurrence of EC signals can produce misleading signatures that resemble those of the target events, such as Weakly Interacting Massive Particles (WIMPs)~\cite{XENONWIMP2019PhysRevD.103.063028,LUXWIMP2020PhysRevD.101.042001,PANDAXWIMPXIA2019193,XMASSWIMPABE201378, DARKSIDEWIMPPhysRevLett.121.081307} and coherent elastic neutrino-nucleus scattering (CE$\nu$NS)~\cite{XENONnTJCAP2020,Aalbers_2016,LUXZEPLIN2020PRD}. Furthermore, to accurately measure the two-neutrino double-electron capture in $^{124}\mathrm{Xe}$, an essential background contribution arises from $^{125}\mathrm{I}$ EC, whose decay peak closely overlaps with that of the two-neutrino double-electron capture peak~\cite{AprilePRC2022}. Because of their significance, the $^{125}\mathrm{I}$ EC fractions were examined experimentally as recently as 2022~\cite{KaurJLTF2022}.

Considering the increasing interest in the field and its broad applicability, we re-examine the EC formalism and the calculations of capture fractions. For the bound electron wave functions, we employ the Dirac-Hartree-Fock-Slater (DHFS) self-consistent framework, which accounts for electron screening and correlations in the atomic structure description. To substantiate the efficacy of the DHFS approach, we systematically compare the binding energies and Coulomb amplitudes with prior theoretical models and experimental data. To calculate the electron capture fractions, we account for several critical atomic effects, including the overlap and exchange corrections, shake-up, and shake-off phenomena. Some of these atomic effects have been proven important in nuclear weak interaction processes, especially for low energy transitions \cite{HarstonPRA1992,MougeotPRA2012,MougeotPRA2014,NitescuPRC2023}.

It is not the scope here to review the rich history of EC calculations. For a comprehensive overview, one can check \cite{BambynekRMP1976}. In what follows, we highlight just the notable differences between our model and the most recent calculations \cite{MougeotARI2018,MougeotARI2019,MougetasrXiv2021}. Firstly, the electron binding energies entering the EC rate are obtained from different atomic structure descriptions. While we use the DHFS self-consistent framework, recent models force the convergence of these energies to a particular set of values for each atomic number. Our treatment leads to better agreement with experimental values for the binding energies, providing a more accurate description of the atomic structure, especially for inner shells, from which EC is most probable. Secondly, our model calculates the overlap and exchange corrections, as well as the shake-up and shake-off effects exactly, by considering the final states of the electrons based on the configuration of the final atom. In contrast, previous models compute the overlaps by deriving the final atom orbitals from the ones of the initial atom using first-order perturbation theory, which may introduce additional uncertainties in the results.

A key feature, that distinguishes our model from previous ones, is the the use of a more refined energy balance of the EC process, with atomic masses which avoids approximating the total electron binding energies of the atomic systems and allows a more precise determination of the neutrino energy. The novel aspect entails extensive computations of the structure of both the excited and ground states of the final atomic system. However, the endeavor is justified considering how sensitive the decay rate for the EC process is to the neutrino energy. Our findings indicate that the refined energetics yields better agreement with the experimental values for the electron capture fractions. The advancements are most pronounced in the low energy transitions, where the limitations of the earlier models' approximations on the total binding energy become noticeable.

Finally, we include in the evaluation of uncertainties the effects related to the atomic structure calculation. This is important as it affects the determination of the emitted neutrino energy and the resulting decay rate, as discussed above. To address this aspect, we introduce an associated uncertainty for the atomic relaxation energies, providing a more comprehensive assessment of overall uncertainty in electron capture observables. Additionally, our method using pseudo-experiments ensures 68\% coverage, in contrast to the ad-hoc assessment of uncertainties in previous models.

The present paper is structured as follows. In Section~\ref{sec:ECformalism}, we provide the EC formalism, encompassing the key equations for the EC decay rates, the novelty of the energetics of the process, and the essential atomic effects. Subsequently, Section~\ref{sec:ElectraonWaveFunction} outlines the DHFS self-consistent method for determining the electron wave functions and the atomic potential for solving the Dirac equation. Herein, we calculate the binding energies, Coulomb amplitudes, and total atomic binding energies, which we compare to prior calculations and experimental data. Section~\ref{sec:Results} is devoted to presenting and analyzing our outcomes for the EC fractions. Concluding remarks are presented in Section~\ref{sec:Conslusions}.

\section{Electron capture formalism}
\label{sec:ECformalism}

We investigate the electron capture process,
\begin{equation}
	e^-+(A,Z)\rightarrow(A,Z')^*+\nu_e,
\end{equation}
in which the initial nucleus $(A,Z)$ captures one atomic electrons, changing its atomic number by one unit ($Z'=Z-1$) and emitting one neutrino. The transition probability per unit time that the electron capture process occurs from all atomic shells is given by \cite{Behrens1982,BambynekRMP1976,FirestoneBook1999,MougeotARI2018,MougeotARI2019,MougetasrXiv2021}
\begin{equation}
	\lambda=\frac{G_{\beta}^{2}}{2 \pi^{3}}\sum_{\textrm{x}}n_{\textrm{x}}C_{\textrm{x}}F_{\textrm{x}}S_{\textrm{x}}, 
\end{equation} 
where $G_{\beta}$ is the weak interaction coupling constant. The sum runs over all partially filled or closed atomic shells $\textrm{x}$, from which the electron can be captured, with the relative occupation number $n_{\textrm{x}}$. For closed shells $n_{\textrm{x}}=1$ and for partially filled ones $n_\textrm{x}=N_\textrm{x}/(2j_\textrm{x}+1)$, where $N_\textrm{x}$ is the number of electrons, with total angular momentum $j_\textrm{x}$, partially filling the shell. The term $C_{\textrm{x}}$ contains the nuclear matrix elements, and $S_{\textrm{x}}$ is related to the atomic shake-up and shake-off effects. The latter will be detailed in Section~\ref{sec:Shake}. The function $F_{\textrm{x}}$ is given by
\begin{equation}
	F_{\textrm{x}}=\frac{\pi}{2}q^2_{\textrm{x}}\beta^2_{\textrm{x}}B_{\textrm{x}},
\end{equation}
and it corresponds to the integrated Fermi function for $\beta$ decay. Here, $q_\textrm{x}$ is the energy of the emitted neutrino (see Section~\ref{sec:Energetics}), $B_{\textrm{x}}$ includes the overlap and exchange effects (see Section~\ref{sec:Exchange}), and $\beta_{\textrm{x}}$ is the so-called Coulomb amplitudes related to the power expansion of the wave function for the captured electron (see Section~\ref{sec:CoulombAmplitudes}).  

For ($L-1$)-unique-forbidden electron capture process, one can write the transition probability from all atomic shells in a simplified form,
\begin{equation}
	\label{eq:decay_rate_CoulAmp}
	\lambda=M^2_L\frac{(2 L-2) ! !}{(2 L-1) ! !}\sum_{\textrm{x}}\frac{n_{\textrm{x}} p_{\textrm{x}}^{2\left(k_\textrm{x}-1\right)} q_{\textrm{x}}^{2\left(L-k_\textrm{x}+1\right)}\beta^2_{\textrm{x}}B_{\textrm{x}}S_{\textrm{x}}}{\left(2 k_\textrm{x}-1\right) !\left[2\left(L-k_\textrm{x}\right)+1\right] !},
\end{equation}
where $M_L$ contains the nuclear matrix element, $L$ is the electron capture transition angular momentum, and the bound electron linear momentum is given by 
\begin{equation}
	p_{\textrm{x}}=\sqrt{m_e^2-W^2_{\textrm{x}}},
\end{equation}
where $m_e$ is the electron rest-mass energy. Here, $W_{\textrm{x}}=m_e-\left|E_{\textrm{x}}\right|$ is the total energy of the captured electron in the initial atom, with the binding energy $E_{\textrm{x}}$.  We are using the system of units such that $\hbar=c=1$. The positive integer number $k_{\textrm{x}}=\left|\kappa\right|$, where $\kappa$ is the relativistic quantum  of the shell $\textrm{x}$, from where the electron is captured. To keep the notation simple, we stick with the X-ray notation for the shells, i.e., $\textrm{x}=K$, $L_{1}$, $L_{2}$, $L_{3}$, $M_{1}$, $M_{2}$, $\ldots$. In the following sections, it will be advantageous to use the spectroscopic notation, in which the atomic shells can be identified by $(n \kappa)$ or $(n\ell_j)$. Here, $n$, $\ell$, and $j$ are the principle, the orbital angular momentum, and the total angular momentum quantum numbers of the subshell.   

For allowed transitions, $k_{\textrm{x}}=1$, so, $K$, $L_1$, $L_2$, $M_1$, $M_2$, $\ldots$ atomic shells contribute to the total decay rate,
\begin{equation}
	\lambda=\lambda_K+\lambda_{L_1}+\lambda_{L_2}+\ldots
\end{equation}
where the sum stops at the last occupied shell with $k_{\textrm{x}}=1$ of the initial atom. Experimentally, one can measure the electron capture shell or subshell ratios, e.g., $L_1/K$ capture ratio, which is derived from Eq.~\ref{eq:decay_rate_CoulAmp} as
\begin{equation}
	\label{eq:lambda_l1_over_k_allowed}
	\frac{\lambda_{L_1}}{\lambda_K}=\frac{n_{L_1}q^2_{L_1}\beta^2_{L_1}B_{L_1}S_{L_1}}{n_Kq^2_K\beta^2_{K}B_{K}S_{K}}.
\end{equation}   
The identification of other ratios for allowed electron capture is straightforward.

In the case of the first unique forbidden electron capture, the total decay rate can be written as a sum of two contributions,
\begin{equation}
	\lambda=\lambda^{(1)}+\lambda^{(2)}
\end{equation}
where $\lambda^{(1)}$ includes a sum over the atomic shells with $k_{\textrm{x}}=1$ (the same as in the allowed electron capture case), and $\lambda^{(2)}$ includes a sum over the atomic shells with $k_{\textrm{x}}=2$ ($L_3$, $M_3$, $M_4$, $\ldots$). In the case of the first unique forbidden electron capture, the $L_1/K$ and $L_3/K$ ratios are derived from Eq.~\ref{eq:decay_rate_CoulAmp} as,
\begin{equation}
	\label{eq:lambda_l1_over_k_fuf}
	\frac{\lambda_{L_1}}{\lambda_K}=\frac{n_{L_1}q^4_{L_1}\beta^2_{L_1}B_{L_1}S_{L_1}}{n_Kq^4_K\beta^2_{K}B_{K}S_{K}},
\end{equation}
and   
\begin{equation}
	\label{eq:lambda_l3_over_k_fuf}
	\frac{\lambda_{L_3}}{\lambda_K}=\frac{n_{L_3}p^2_{L_3}q^2_{L_3}\beta^2_{L_3}B_{L_3}S_{L_3}}{n_Kq^4_K\beta^2_{K}B_{K}S_{K}},
\end{equation}
respectively. Other ratios can be derived in an analogous manner.


\subsection{Energetics}
\label{sec:Energetics}
We adopt the following energetic balance for the electron capture process,
\begin{equation}
	Q=q_{\textrm{x}}+\mathcal{M}^{(*)}_{\textrm{x}}(A,Z')-\mathcal{M}_\textrm{gs}(A,Z')
\end{equation}
where $Q=\mathcal{M}_\textrm{gs}(A,Z)-\mathcal{M}_\textrm{gs}(A,Z')$ is the atomic mass difference between the initial and final systems in ground states, and the mass difference in the r.h.s. is the energy emitted through X-rays and $\gamma$-rays by the final system. The atomic mass, $\mathcal{M}^{(*)}_{\textrm{x}}(A,Z')$, corresponds to an atomic and possible nuclear excited state of the final system (with a hole in shell $\textrm{x}$ and the nuclear excited state with energy $R_\gamma$).   

Taking into account that the atomic mass of the final excited system can be written in terms of the mass of the final nucleus in ground state, $M_f$, and the total electron binding energy, $B(Z')$, as \cite{AME2020}
\begin{equation}
	\label{eq:AtomicMass}
	\mathcal{M}^{(*)}_{\textrm{x}}(A,Z')=M_f+R_\gamma+Z'm_e-B_\textrm{x}(Z'),
\end{equation}
the emitted neutrino energy is given by

\begin{align}
	\label{eq:NewNeutrinoEnergy}
	\begin{aligned}
		q_{\textrm{x}}&=Q-R_\gamma-\left[B_{\textrm{gs}}(Z')-B_{\textrm{x}}(Z')\right]\\
		&=Q-R_\gamma-R_\textrm{x}. 
	\end{aligned}
\end{align}
One can see that writing $Q$ in explicit form, $Q=\Delta M_{if}+m_e+\left[B_{\textrm{gs}}(Z')-B_{\textrm{gs}}(Z)\right]$, where $\Delta M_{if}$ is the difference between the ground states energies of the initial and final nucleus, and making the approximation $\left[B_{\textrm{x}}(Z')-B_{\textrm{gs}}(Z)\right]\approx-\left|E_\textrm{x}\right|$ in Eq.~\ref{eq:NewNeutrinoEnergy}, we recover the neutrino energy as it was approximated in the previous approaches \cite{BambynekRMP1976,FirestoneBook1999,MougeotARI2018,MougeotARI2019,MougetasrXiv2021}

\begin{align}
	\label{eq:OldNeutrinoEnergy}
	\begin{aligned}
	q_{\textrm{x}}	&=\Delta M_{if}-R_{\gamma}+m_e-\left|E_\textrm{x}\right|\\
					&=W_0+W_{\textrm{x}},
	\end{aligned}
\end{align}
where $W_0=\Delta M_{if}-R_{\gamma}$ it is usually called the electron capture transition energy. The neutrino energy in Eq.~\ref{eq:OldNeutrinoEnergy} neglects the total change in electron binding energy between initial and final atoms and the rearrangement energy of the captured electron. In our approach, we consider these quantities through the atomic relaxation energy, $R_\textrm{x}$. We expect the difference between  Eq.~\ref{eq:OldNeutrinoEnergy} and Eq.~\ref{eq:NewNeutrinoEnergy} to play a significant role in low $Q-R_{\gamma}$ transitions. In both determinations of the neutrino energy, the nuclear recoil energy is neglected, as its largest value is only $57$ eV in the EC of $^{7}$Be \cite{BambynekRMP1976}.

Working with atomic masses offers a clear advantage over nuclear masses. Nuclear masses are typically difficult to measure, while $Q$-values with small uncertainties can be obtained from the mass excesses provided in \cite{AME2020}. Furthermore, nuclear excitation energies $R_\gamma$ for low-lying states that are usually populated in the final nucleus by electron capture are available with high precision \cite{livechart}. A precise determination of the neutrino energy depends on the accuracy of the atomic de-excitation energy, $R_\textrm{x}$, calculation. We use the DHFS self-consistent framework for this purpose, and the results and associated uncertainties are discussed in Section~\ref{sec:TotalElectronBindingEnergy}.

\subsection{Overlap and exchange corrections}
\label{sec:Exchange}

When the nucleus captures an electron, the other electrons do not participate such that they could be considered spectators and, in the first approximation, their contribution could be neglected. But due to the change of the nuclear charge by one unit, the wave functions of the spectator electrons change such that the overlap between their initial state $\ket{(m, \kappa)}$ and final state $\ket{(m, \kappa)^{\prime}}$ is imperfect $\bra{(m, \kappa)^{\prime}}\ket{(m, \kappa)}\neq1$. All those imperfections have a non-negligible effect which requires an overlap correction.

After the capture of the electron from a specific shell, a hole can be observed in that respective shell. This is called a direct capture. But due to the indistinguishability of the electron, the process can have different behaviour. First, an electron from another shell than the one where the hole is observed is captured. At the same time, an electron from the shell where the hole is observed is promoted in the place of the initial one. Then, from an experimental point of view, the hole is observed in the same shell, making those two processes impossible to distinguish. This is the so-called exchange effect.

These two effects have been taken into account first by Bahcall \cite{BahcallPRL1962, BahcallPR1963, BahcallPR1963B, BahcallNP1965}, who included in the EC transition probability a term considering these exchange and overlap corrections, defined as:

\begin{equation}
	B_{n \kappa}=\left|\frac{b_{n \kappa}}{\beta_{n \kappa}}\right|^{2}
\end{equation}

Here, the capture amplitude $b_{n \kappa}$ include contributions due to these effects from only the first three orbitals with $\kappa=-1$, namely $1s_{1/2}$, $2s_{1/2}$, $3s_{1/2}$, assumes a complete set of states for the other orbitals and use the the closer property for the summation over the continuous states. 

Later an extension was made by Vatai \cite{VataiNPA1970} who extended the formalism by considering in addition the $4s_{1/2}$ orbital and the overlap correction for every subshell.

The explicit expressions of the $b_{nk}$ capture amplitudes used by Bahcall and Vatai can be found in the above mentioned references of their works. 
We note that Vatai didn't include the shake-off and shake-up corrections in his calculations, while Bahcall, does consider it indirectly by using the closure approximation.

Recently there was a generalization of both Bahcall's and Vatai's exchange and overlap corrections in \cite{MougeotARI2018}. We employ the generalization for Vatai's approach as follows \cite{MougeotARI2018}:

\begin{eqnarray}
	b_{n \kappa}=&&\left[\prod_{m, \mu}\bra{(m, \mu)^{\prime}}\ket{(m, \mu)}^{n_{m \mu}}\right] \bra{(n, \kappa)^{\prime}}\ket{(n, \kappa)}^{-\frac{1}{2|\kappa|}}\nonumber\\
	&&\times\left[\beta_{n \kappa}-\sum_{\substack{m \neq n}} \beta_{m \kappa} \frac{\bra{(m, \kappa)^{\prime}}\ket{(n, \kappa)}}{\bra{(m, \kappa)^{\prime}}\ket{(m, \kappa)}}\right]
\end{eqnarray}

We consider the shake-up and shake-off effects in the next section.

\subsection{Shake-up and shake-off effects}
\label{sec:Shake}
The shake-up effect considers the probability that a spectator electron will be promoted to a vacant upper shell during the decay. At the same time, the shake-off considers the possibility of such a spectator electron being ejected into continuous. Those effects generate another hole in the atomic shell beside the one the electron capture produces. The $S_{\textrm{x}}$ term factorises the case with no second hole and the probability of having a second hole in each shell. Thus it is defined as \cite{MougeotARI2018}:
\begin{equation}
	S_{\textrm{x}}=1+\sum_{m, \mu}P_{m \mu}
\end{equation}

A spectator electron has three possible behaviours during the decay: shake-up, shake-off or remaining in the initial electronic cloud. One could calculate the total shaking probability by subtracting the probability of staying in the atomic shell from unity. Thus, the probability of an electron from the $(m, \kappa )$ shell to undergo the shake-up or the shake-off process is represented as \cite{MougeotARI2018}:

\begin{eqnarray}
	P_{m \mu}=&&1-|\bra{(m, \mu)^{\prime}}\ket{(m, \mu)}|^{2 n_{m\mu}}\nonumber\\
	&&-\sum_{l \neq m} n_{l \mu}^{\prime} n_{m \mu}|\bra{(l, \mu)^{\prime}}\ket{(m, \mu)}|^2
\end{eqnarray}

We emphasize that in our computation of the overlap and exchange corrections and shake-up and shake-off effects as presented in Secs. \ref{sec:Exchange} and \ref{sec:Shake}, the final states of the electrons are calculated taking into account the configuration of the final atom (i.e. with a hole in place of the captured electron). Thus each set of states is specific to an electron capture from a particular shell. For example, the $S_{\textrm{x}}$ term has different values for capture from different shells. In contrast, the model proposed in \cite{MougeotARI2018,MougeotARI2019} employs an approximation for the overlaps of the form $\braket{(m, \kappa)^{\prime}}{(n, \kappa)}$, where the final atom orbitals are computed from the initial atom orbitals using first-order perturbation theory.

\section{Atomic bound states}
\label{sec:ElectraonWaveFunction}

A precise description of the orbital electron capture process requires good knowledge of the bound states wave functions for electrons in the initial atom's ground state configuration and the final atom's excited state configurations. If we assume a central field, $V(r)$, for the atomic system, the electron bound states relativistic wave functions can be separated as \cite{RoseBook1961},
\begin{equation}
\psi_{n \kappa m}(\boldsymbol{r})=\left(\begin{array}{c}
g_{n \kappa}(r) \Omega_{\kappa,m} (\hat{\boldsymbol{r}}) \\
i f_{n \kappa}(r) \Omega_{-\kappa,m} (\hat{\boldsymbol{r}})
\end{array}\right),
\end{equation} 
where $\Omega_{\kappa,m}$ are the spherical spinors \cite{RoseBook1995,VarshalovichBook1988} and $\boldsymbol{r}$ stands for the position vector of the electron, $r=\left|\boldsymbol{r}\right|$ and $\hat{\boldsymbol{r}}=\boldsymbol{r}/r$.

The functions $g_{n \kappa}(r)$ and $f_{n \kappa}(r)$ are the large- and small-component radial wave functions, respectively, and they obey the following system of coupled differential equations,
\begin{align}
\begin{aligned}
\label{eq:radialEquations}
\left(\frac{d}{dr}+\frac{\kappa+1}{r}\right)g_{n\kappa}-(W_{n\kappa}-V(r)+m_e)f_{n\kappa}&=0,\\
\left(\frac{d}{dr}-\frac{\kappa-1}{r}\right)f_{n\kappa}+(W_{n\kappa}-V(r)-m_e)g_{n\kappa}&=0.
\end{aligned}
\end{align}
Here the relativistic quantum number, $\kappa$, takes positive and negative integer values and identifies both the total angular momentum, $j$, and the orbital angular momentum, $\ell$, by
\begin{eqnarray}
j=\left|\kappa\right|-1/2, \hspace{0.2cm} \ell=\begin{cases}
\kappa  &\quad \text{if } \kappa>0,\\
\left|\kappa\right|-1  &\quad \text{if } \kappa<0.\\
\end{cases}
\end{eqnarray}
Hence a $(2j+1)-$degenerate atomic bound state can be identified by either ($n\kappa$) or ($n\ell_j$) notation. In the spectroscopic notation, the latter is given by $n\ell_j=1s_{1/2}, 2p_{1/2}, 2p_{3/2},\ldots$.

\begin{figure}[t]
	\centering{
		\includegraphics[width=0.48\textwidth]{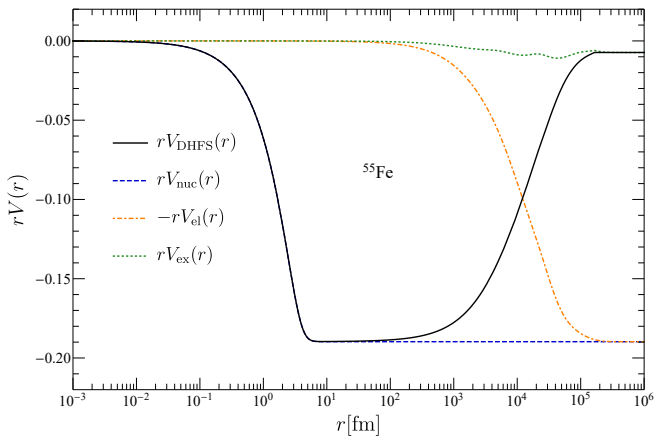}
	}
	\caption{\label{fig:potential} The DHFS potential (black) for neutral atom $^{55}$Fe in ground state electronic configuration multiplied by the radius, along with its components: nuclear potential (blue dashed), electronic potential (orange dot-dashed) and exchange potential (green dotted).}
\end{figure}

For the calculation of the atomic structure, we employed the \textsc{RADIAL} subroutine package \cite{SalvatCPC2019}, which includes the program \textsc{DHFS.f}. The latter solves the DHFS equations for the ground state or excited states of neutral atoms and positive ions with $N_e$ bound electrons and $Z_p$ protons in the nucleus. The EC calculations involve only neutral atoms, i.e., $N_e=Z_p$. Here, $Z_p$ can be either the atomic number of the initial atom ($Z$) in the g.s. or the atomic number of the final atom ($Z'$) in an excited state (with a hole from where the electron is captured). The atomic potential is composed as \cite{SalvatCPC2019}:
\begin{equation}
V_{\mathrm{DHFS}}(r)=V_{\mathrm{nuc}}(r)+V_{\mathrm{el}}(r)+V_{\mathrm{ex}}(r),
\end{equation}
where $V_{\mathrm{nuc}}(r)$, $V_{\mathrm{el}}$ and $V_{\mathrm{ex}}(r)$ are respectively the nuclear, electronic and exchange potential. The nuclear potential is generated by a realistic Fermi proton charge distribution $\rho_p(r)$ \cite{HahnPR1956}, and the electronic potential by the atomic electron cloud distribution $\rho(r)$. The exchange term in the DHFS potential is simplified due to Slater's approximation \cite{SlaterPR1951}, which takes the exchange potential proportional to the atomic electron cloud distribution to the power $1/3$. The Slater exchange potential is inadequate for large distances from the atom, where the electron density is very small, and, hence, the obtained DHFS potential does not respect the correct asymptotic condition when $r\rightarrow\infty$. The drawback is solved by the Latter's tail correction \cite{LatterPR1955}, and the local exchange potential with the correct asymptotic behavior is given by, 
\begin{eqnarray}
\label{eq:LatterTail}
V_\text{ex}(r)=
\begin{cases} 
-\frac{3}{2}\alpha	\left(\frac{3}{\pi}\right)^{1/3}\left[\rho(r)\right]^{1/3} & r < r_{\text{Latter}}, \\\\
-\frac{\alpha(Z_p-N_e+1)}{r}-V_{\text{nuc}}(r)-V_{\text{el}}(r) & r\geq r_{\text{Latter}}. 
\end{cases}
\end{eqnarray} 
where $\alpha$ is the fine-structure constant and $r_{\text{Latter}}$ is the radius where the total DHFS potential starts to deviate from the correct asymptotic value.

The electron density, $\rho(r)$, is calculated self-consistently \cite{LibermanCPC1971,LibermanPR1965}. The procedure starts with an approximate electron density obtained as the Moli\`ere parametrization of the Thomas-Fermi potential \cite{MoliereZNA1947}. Then, the electron density is renewed iteratively from the obtained bound wave functions until the DHFS potential converges within a specified tolerance. We depict in Fig.~\ref{fig:potential} the stabilized DHFS potential along with its components for the neutral atom $^{55}$Fe. One can see that for large radii, the potential respect the correct asymptotic condition, i.e., $rV_{\textrm{DHFS}}(r)\rightarrow-\alpha$. In this case, the cut-off radius, $r_{\text{Latter}}$, is around $1.7\times10^{5}$ fm.

\subsection{Electron binding energies and wave functions}

The choice of the atomic potential generated by the nuclear charge and the atomic electron cloud, $V(r)$, is crucial when solving for electron bound states. 
One can derive an approximation of $V(r)$ by solving the Thomas-Fermi or Thomas-Fermi-Dirac equations \cite{ThomasJCP1954,LatterPR1955,GombasBook1956}. However, if one aims for higher precision the self-consistent Hartree-Fock or Dirac-Hartee-Fock (DHF) method should be employed \cite{HartreeBook1957,GrantAP1970}. In what follows, we shall motivate our choice of using the DHFS self-consistent framework in the context of the atomic electron capture process.

DHFS is a version of the DHF in which $V_\textrm{ex}$ is modified as mentioned in the previous section. The self-consistent DHF method is considered one of the most reliable methods for atomic structure calculations, providing an exact treatment of the exchange potential \cite{DesclauxADNUDT1973}. Other approaches that offer excellent agreement with experimental values for spin-orbit splitting  ionization potential, and total atomic energy are the local density approximation methods \cite{KotochigavaPRA1997}, e.g., relativistic local density approximation (RLDA). However, for inner shells, from where electron capture is most probable, the DHFS self-consistent method provides better agreement with experimental values regarding the binding energies. Interestingly, increasing the sophistication involved in the DHF method does not necessarily lead to an increased agreement with experimental binding energies \cite{LuADNDT1971}.
 
\begin{figure*}[t]
	\centering
	\includegraphics[width=0.75\linewidth]{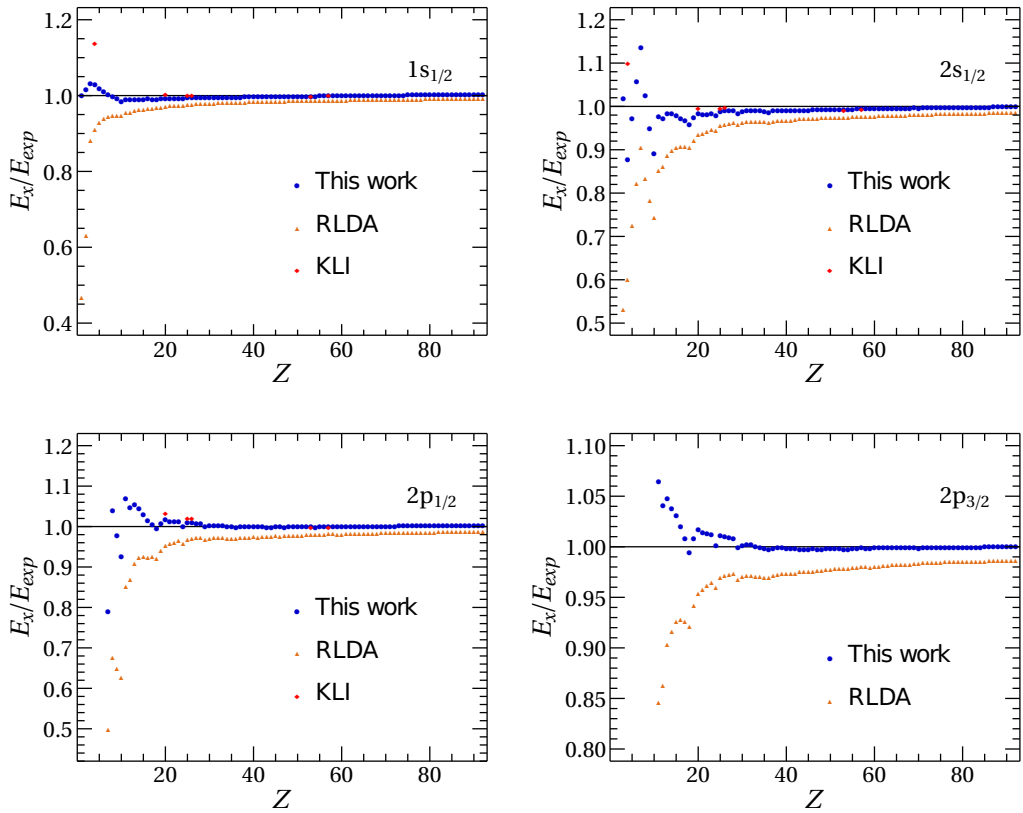}
	\caption{\label{fig:Ebind}The ratios between predicted and experimental binding energies within the relativistic local density approximation framework (orange triangles), the KLI model (red diamonds) and the DHFS self-consistent framework (blue dots). The experimental values are taken from Ref.~\cite{LotzJOSA1970}, the RLDA predictions can be found online \cite{KotochigavaNIST2021} and the KLI predictions are taken from \cite{MougetasrXiv2021}. The results are presented for 1s$_{1/2}$, 2s$_{1/2}$, 3s$_{1/2}$, 3p$_{1/2}$ orbitals of neutral atoms in the ground state with atomic number Z between 1 and 92.}
\end{figure*}

\begin{table*}
	\begin{center}
		\begin{tabular}{c|cccc|cccc|cccc}
			\hline \hline & \multicolumn{4}{|c|}{$E_{1s_{1/2}}$ (eV)} & \multicolumn{4}{|c|}{$E_{2s_{1/2}}$ (eV)} & \multicolumn{4}{|c}{$E_{2p_{1/2}}$ (eV)} \\
			Isotope & RLDA & KLI  & DHFS & EXP & RLDA & KLI  & DHFS & EXP  & RLDA & KLI  & DHFS & EXP  \\
			\hline${ }^{7} \mathrm{Be}$ & $-104.9$ & $-131.0$ & $-118.4$ & $-115$ & $-5.599$ & $-10.259$ & $-8.181$ & $-9.322$ & $-$ & $-$ & $-$ &$-$ \\
			${ }^{41} \mathrm{Ca}$ & $-3929.4$ & $-4052.6$ & $-4015.1$ & $-4041$ & $-412.4$ & $-439.2$ & $-434.1$ & $-441$ & $-336.7$ & $-364.4$ & $-359.0$ & $-353$ \\
			${ }^{54} \mathrm{Mn}$ & $-6397.0$ & $-6552.8$ & $-6510.9$ & $-6544$ & $-740.6$ & $-772.3$ & $-766.4$ & $-775$ & $-635.9$ & $-669.6$ & $-662.9$ & $-656$ \\
			${ }^{55} \mathrm{Fe}$ & $-6963.3$ & $-7126.1$ & $-7083.4$ & $-7117$ & $-816.1$ & $-849.0$ & $-842.9$ & $-851$ & $-705.3$ & $-740.4$ & $-733.5$ & $-726$ \\
			${ }^{125} \mathrm{I}$ & $-32765.9$ & $-33162.8$ & $-33165.0$ & $-33176$ & $-5067.8$ & $-5151.9$ & $-5162.2$ & $-5195$ & $-4761.3$ & $-4852.3$ & $-4857.7$ & $-4858$ \\
			${ }^{138} \mathrm{La}$ & $-38483.4$ & $-38922.4$ & $-38944.2$ & $-38928$ & $-6132.4$ & $-6225.7$ & $-6242.2$ & $-6269$ & $-5790.8$ & $-5891.5$ & $-5902.8$ & $-5894$ \\
			\hline \hline
		\end{tabular}
	\end{center}
	\caption{\label{tab:BindingEnergies} The experimental binding energies (EXP) in comparison with DHFS, KLI and RLDA models described in the text. All binding energies are presented in eV for the inner shells of one light and a few medium and heavy neutral atoms in the ground state.}
\end{table*}

Recent EC calculations have been performed using another framework, included in the program BetaShape (BS), derived from the DHFS method \cite{MougeotARI2018, MougeotARI2019}, using an atomic potential explained in \cite{MougeotPRA2014}. There are two differences compared to our framework, namely, the nuclear potential is considered for a uniformly charged sphere, and an adjustable parameter controls the exchange potential's strength. The latter is used to force convergence of the binding energies to the RLDA predictions for Z=1--92 and to relativistic Dirac-Fock predictions \cite{DesclauxADNUDT1973} for higher atomic numbers. Even more recently \cite{MougetasrXiv2021}, another framework for computing EC observables has been developed, using a self-interaction-corrected model (KLI) \cite{KriegerPRA1992,LiPRA1993} with the exchange-correlation potential from \cite{VoskoCJP1980,VoskoPRB1980}.

To compare the DHFS framework with the RLDA and KLI models, we plot the ratios between their predictions and experimental binding energies for ground-state neutral atoms in the range Z=1--92 in Fig.~\ref{fig:Ebind}. The DHFS method provides ratios closer to unity for the $1s_{1/2}$, $2s_{1/2}$, $2p_{1/2}$, and $2p_{3/2}$ shells. Discrepancies in binding energies for light atoms are due to the mean-field approach used to describe the atomic potential. Experimental values are taken from Ref. \cite{LotzJOSA1970}, RLDA predictions are available online \cite{KotochigavaNIST2021} and KLI predictions are taken from \cite{MougetasrXiv2021}. Our results suggest that the DHFS self-consistent method is a reliable and efficient framework for electron capture calculations. Fig.~\ref{fig:Ebind} also suggests that the forced convergence in BS to the RLDA binding energies is motivated for medium and heavy atoms, but for some light atoms, a more appropriate convergence may be to the experimental binding energies.

Table~\ref{tab:BindingEnergies} presents a comparison between experimental binding energies and the predictions of three different models: RLDA, KLI, and DHFS. The table shows that for medium and heavy atoms, KLI and DHFS models offer comparable precision, while the RLDA model has the worst predictions for the inner shells, from where electron capture is most probable. Regarding the light atom (actually $^{7}$Be is the lightest that undergoes electron capture), it seems that the best approach is the DHFS framework.

We mention that this comparison is valid in the case when the neutrino energy is approximated as in Eq.~\ref{eq:OldNeutrinoEnergy}, but even with the improved energetics developed in Eq.~\ref{eq:NewNeutrinoEnergy}, we need the electron binding energy, $E_\textrm{x}$, to determine the momentum of the captured electron. 

\begin{figure}[b]
	\centering
	\includegraphics[width=\linewidth]{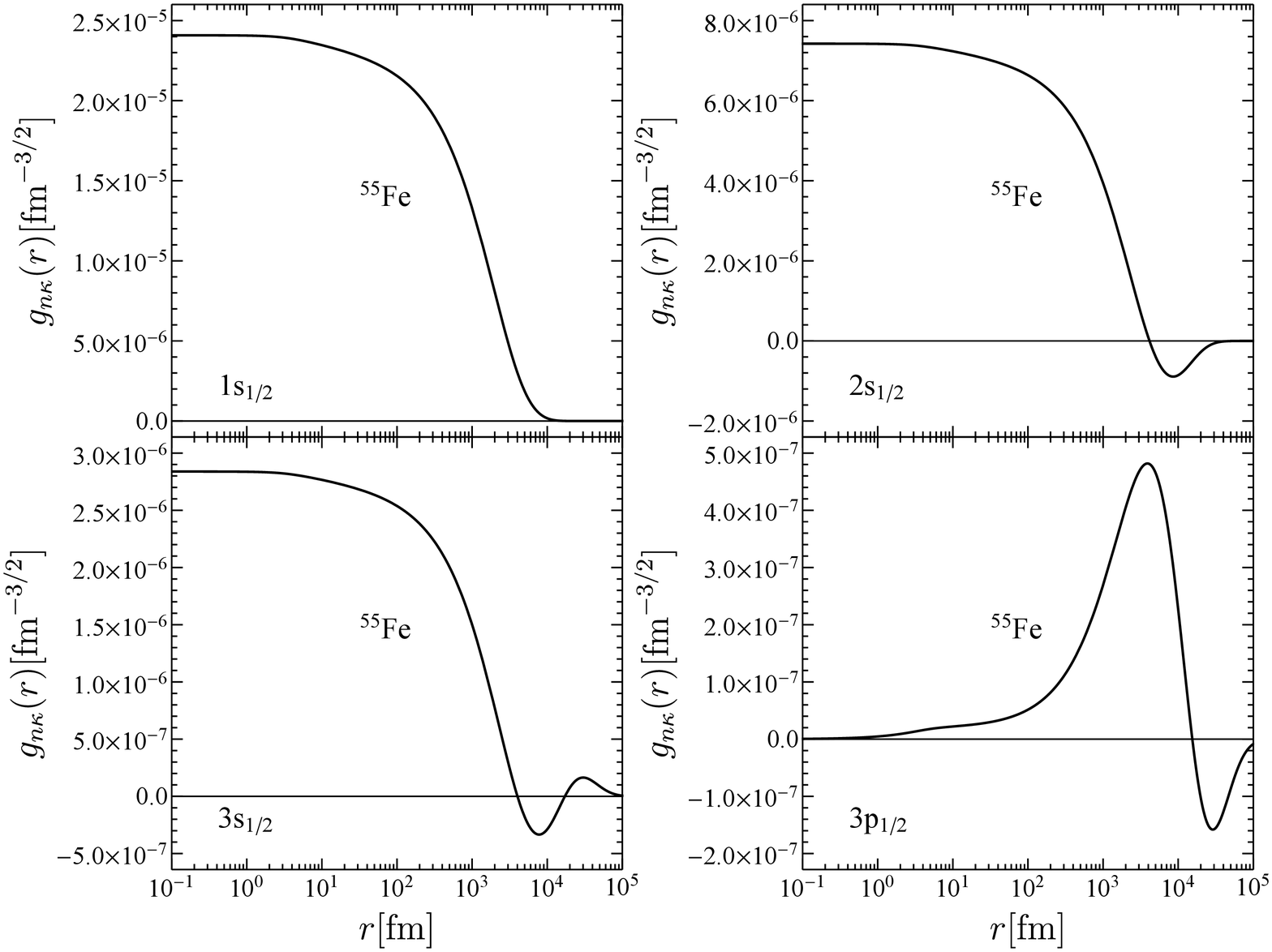}
	\caption{Large-component radial wave functions for the 1s$_{1/2}$, 2s$_{1/2}$, 3s$_{1/2}$, 3p$_{1/2}$ orbitals of ground state neutral atom $^{55}$Fe.}
	\label{fig:gWF}
\end{figure}
\begin{figure}[b]
	\centering
	\includegraphics[width=\linewidth]{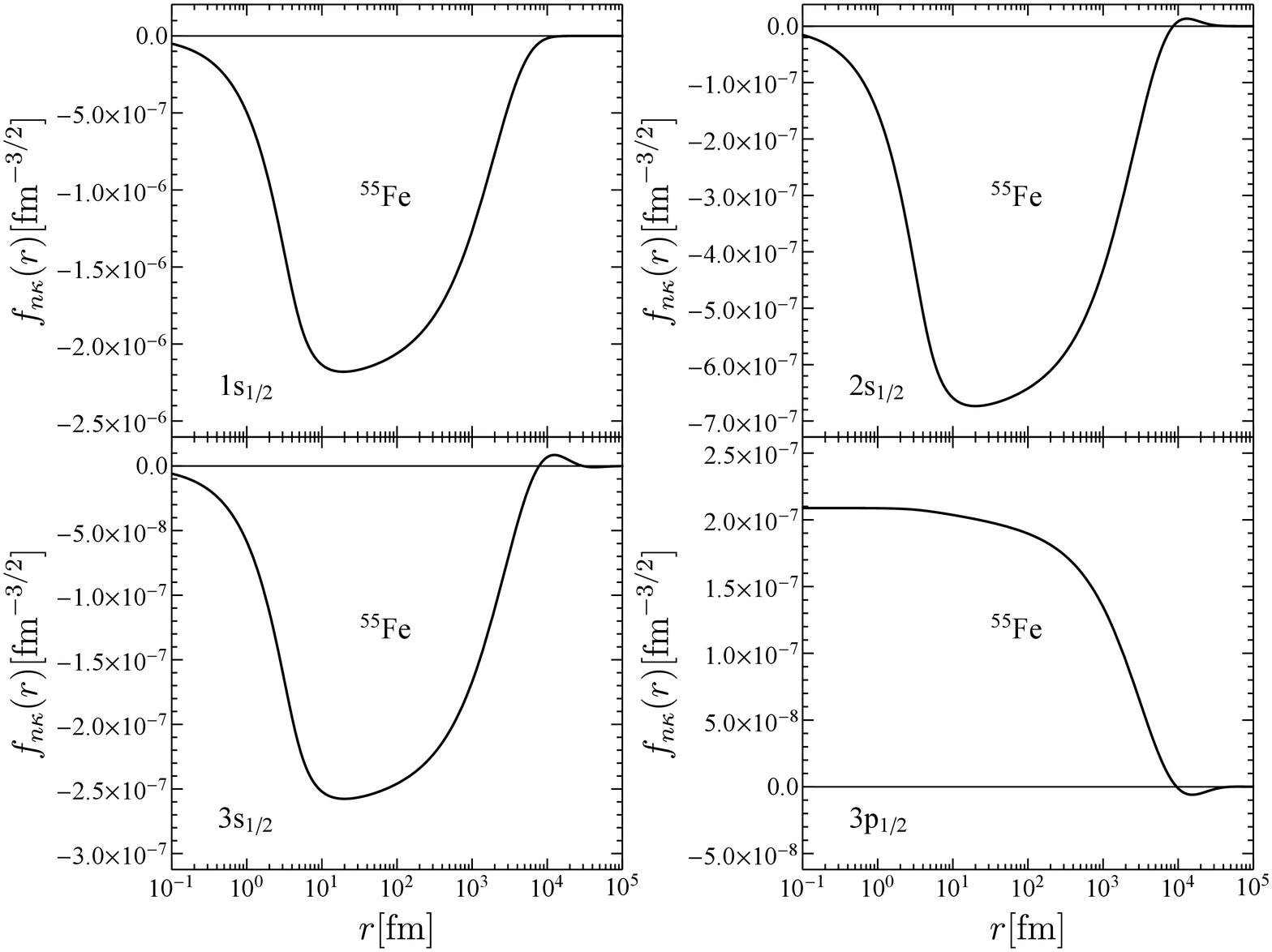}
	\caption{Same as Fig.~\ref{fig:gWF} but for small-component radial wave functions.}
	\label{fig:fWF}
\end{figure}

The accuracy of the bound wave functions also affects the capture probabilities, but the decay rate, which is highly dependent on the transition energetics and the neutrino energy, is more crucial \cite{MougeotARI2019}. Thus, even if the DHFS wave functions are less precise than those of the RLDA or KLI models, the accuracy of the atomic energies compensates for this drawback. To provide a comprehensive view, we include the large- and small-component radial wave functions for the $1s_{1/2}$, $2s_{1/2}$, $3s_{1/2}$, and $3p_{1/2}$ orbitals of the ground state neutral atom $^{55}$Fe in Figs.~\ref{fig:gWF} and \ref{fig:fWF}, respectively. It should be noted that the bound states are orthonormal, i.e.,
\begin{equation}
\int_{0}^{\infty}\left[g_{n\kappa}(r)g_{m\mu}(r)+f_{n\kappa}(r)f_{m\mu}(r)\right]r^2dr=\delta_{nm}\delta_{\kappa\mu}.
\end{equation}

\begin{figure*}
	\centering
	\includegraphics[width=0.8\linewidth]{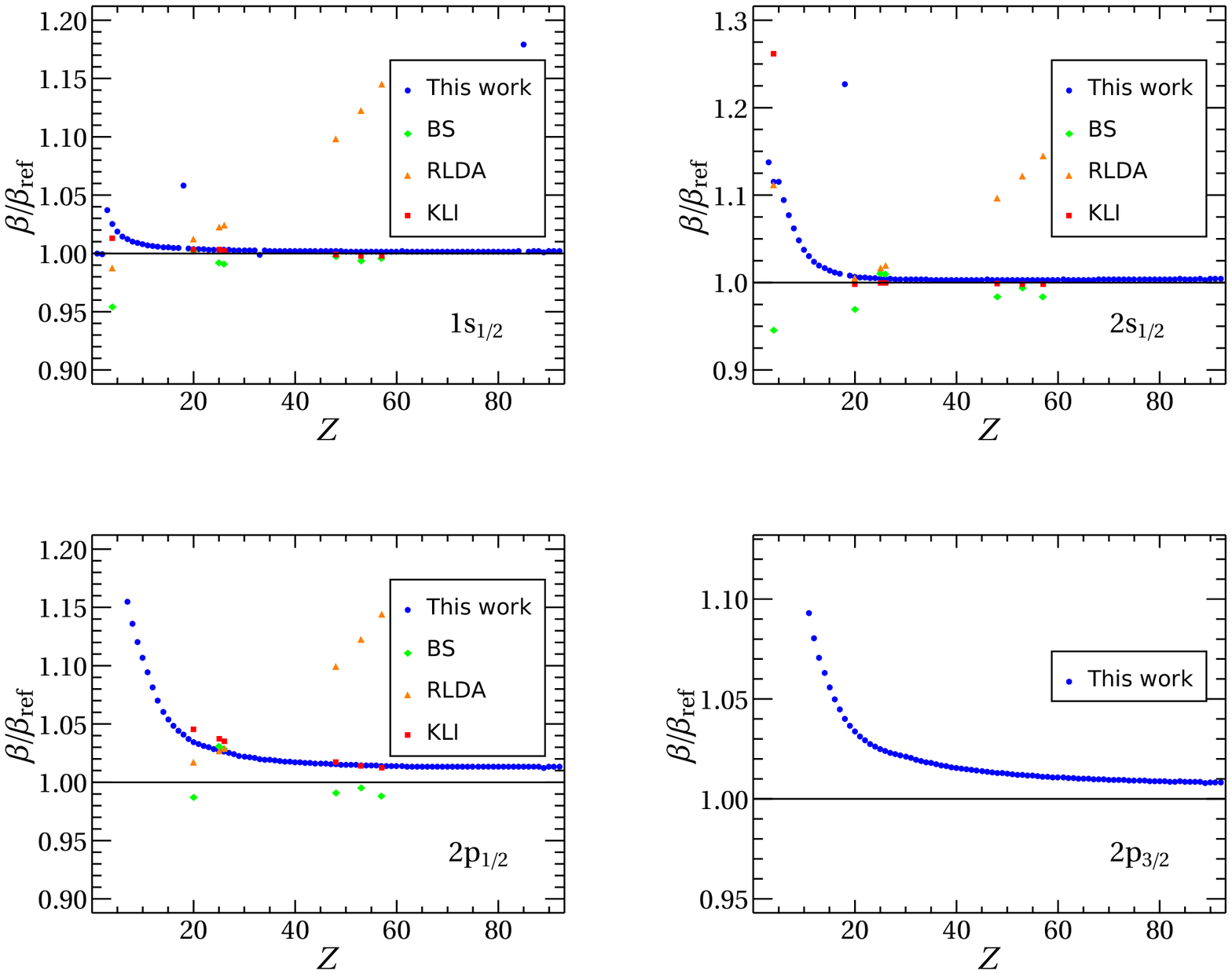}
	\caption{The ratios between the more recently computed Coulomb amplitudes and the reference considering BS (green diamond), RLDA (orange triangles), KLI (red square) and DHFS self-consistent (blue dots). The reference values are given in Ref.~\cite{BambynekRMP1976}, and the BS, RLDA, and KLI results are taken from Ref.~\cite{MougetasrXiv2021}. The ratios are presented for $1s_{1/2}$, $2s_{1/2}$, $3s_{1/2}$, and $3p_{1/2}$ orbitals for ground state atoms having the atomic number Z between 1 and 92.}
	\label{fig:CAmp}
\end{figure*}

\subsection{Coulomb Amplitudes}
\label{sec:CoulombAmplitudes}

The Coulomb amplitude $\beta_{n\kappa}$ is needed in calculating the decay rate. It originates as a constant in the power series of the radial wave function. In Ref.~\cite{Behrens1982}, the expansion series is introduced as: 

\begin{equation}
	\left\{\begin{array}{l}
		g_{n \kappa}(r) \\
		f_{n \kappa}(r)
	\end{array}\right\}=\beta_{n \kappa} \frac{\left(p_{n \kappa} r\right)^{\left|\kappa\right|-1}}{(2\left|\kappa\right|-1) ! !} \sum_{j=0}^{\infty}\left\{\begin{array}{l}
		b_{j} \\
		a_{j}
	\end{array}\right\} r^{j}.
\end{equation}

The $a_j$ and $b_j$ expansion coefficients are defined with a recurrence series

\begin{equation}
	\begin{aligned}
		&(j+\left|\kappa\right|-\kappa) a_{j}=-\left(W_{n\kappa}-m_{e}\right) b_{j-1}+\sum_{m=0}^{j-1} v_{m} b_{j-1-m} \\
		&(j+\left|\kappa\right|+\kappa) b_{j}=\left(W_{n\kappa}+m_{e}\right) a_{j-1}-\sum_{m=0}^{j-1} v_{m} a_{j-1-m}
	\end{aligned}
\end{equation}

with

\begin{equation}
	\begin{array}{ll}
		a_{0}=1, & b_{0}=0 \text { if } \kappa>0 \\
		a_{0}=0, & b_{0}=1 \text { if } \kappa<0
	\end{array}
\end{equation}

The  $v_m$ coefficients describe the atomic potential as a power series:

\begin{equation}
	V_{\mathrm{DHFS}}(r)=\sum_{m=0}^{\infty} v_{m} r^{m}
\end{equation}

The Coulomb amplitude carries a significant quantity of information about the electron state. Thus we compare its behavior depending on the used model, as seen in Fig.~\ref{fig:CAmp} for neutral atoms in the ground-state with the atomic number in the interval Z=1--92. We used as reference the values from \cite{BambynekRMP1976}, which, to our knowledge, provides the only comprehensive systematic analysis of the Coulomb amplitudes across a wide range of atomic numbers. Moreover, the DHF method used in \cite{BambynekRMP1976} treats the exchange potential exactly. Consequently, one can evaluate the impact of the local exchange potential as function of the atomic number. 

The dominant contributions to the decay rate arise from electron captures in the $1s_{1/2}$ and $2s_{1/2}$ orbitals. As can be seen from Fig.~\ref{fig:CAmp}, our model is expected to achieve precision below 1\% for atomic numbers $Z>20$. In contrast, the BS and RLDA models exhibit larger fluctuations across the entire range of atomic numbers. Although there are slight differences (less than 2\%) in our model's predictions for other orbitals in medium and heavy elements, the overall agreement is still maintained. The discrepancies observed for lower atomic numbers can be attributed to the use of the Slater approximation in our model's exchange potential. However, when comparing our DHFS self-consistent calculations with the KLI calculations, which provide a more complete description of the exchange-correlation potential, the results exhibit similar trends. This indicates that there is no significant loss in the description of the bound wave functions. It should be noted that a few data points deviate from the general trend for $Z=18$ and $Z=85$ for the $1s_{1/2}$ orbital, as well as for $Z=18$ for the $2s_{1/2}$ orbital. These discrepancies may be attributed to potential misprints in Ref.~\cite{BambynekRMP1976}.

\subsection{Total electron binding energy}
\label{sec:TotalElectronBindingEnergy}

\begin{figure}[h]
	\centering
	\includegraphics[width=0.49\textwidth]{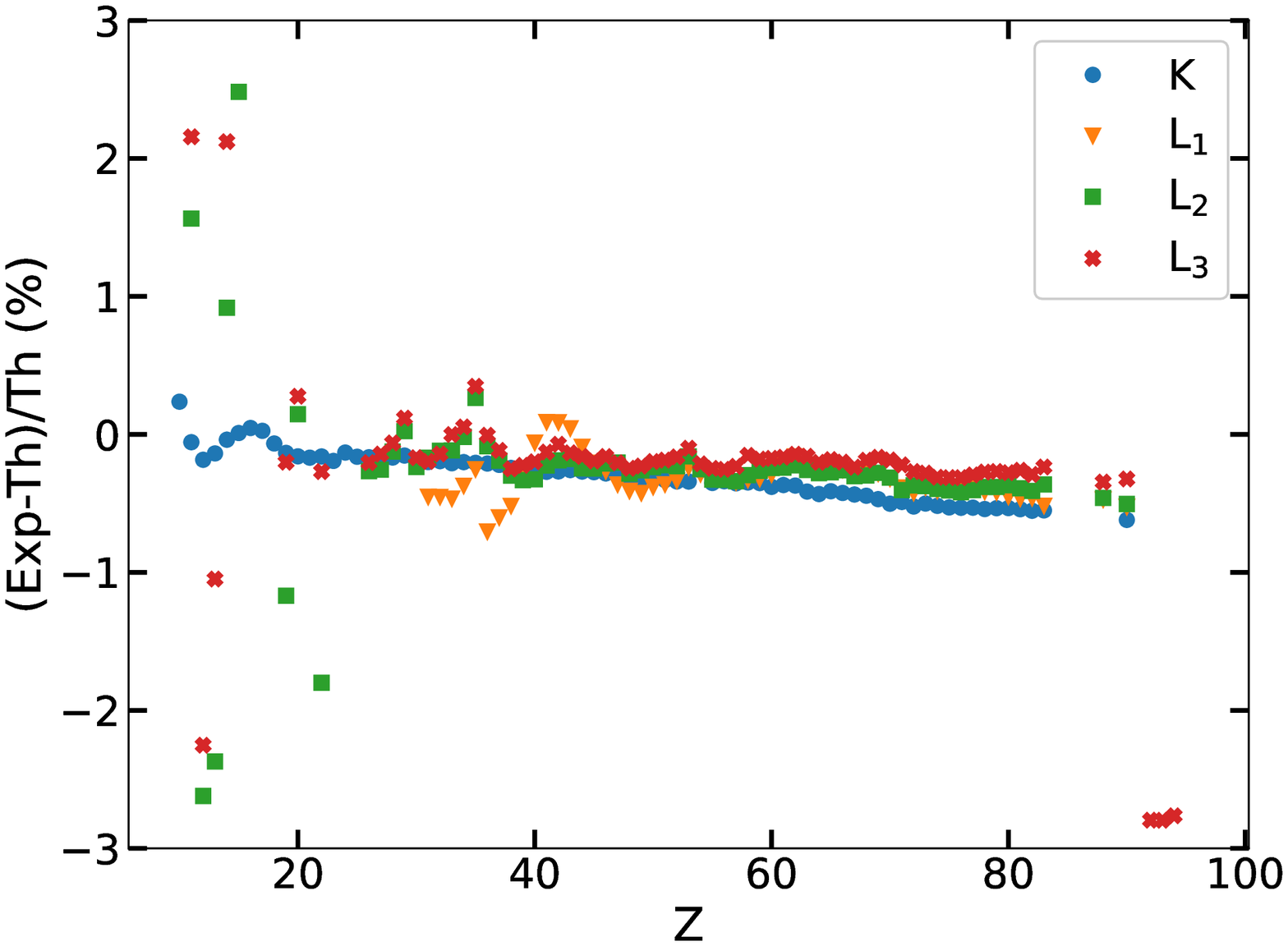}
	\caption{{\small Relative difference between experimental and theoretical values of the atomic relaxation energy as defined in Eq.~\ref{eq:EnergyDepositWithBindingEnergy}, as function of the atomic number. Experimental values are taken from~\cite{DeslattesRMP2003}.}}
	\label{fig:edge_energies_dhfs}
\end{figure}

We turn our attention to the refinements in the energetics done in Section~\ref{sec:Energetics}. The neutrino energy depends on the atomic relaxation energy,
\begin{equation}
	\label{eq:EnergyDepositWithBindingEnergy}
	R_\textrm{x}=B_{\textrm{gs}}(Z')-B_{\textrm{x}}(Z')
\end{equation}
which in our sign convention is the difference in the total electron binding energy of the ground state and excited state final atom. The total electron binding energy is calculated within the program \textsc{DHFS.f}, as the expectation value of the Hamiltonian with the atomic wave function, constructed as a Slater determinant of individual electron wave functions (see Supplementary Material of \cite{SalvatCPC2019}). Despite not including Breit interaction energy, QED corrections, Auger shift, and other factors (see \cite{DeslattesRMP2003}), the DHFS self-consistent framework provides an appropriate level of precision for atomic relaxation energy, given the experimental precision of electron capture. Fig.~\ref{fig:edge_energies_dhfs} shows the relative difference between experimental and computed values of relaxation energy as a function of atomic number, demonstrating that the DHFS self-consistent framework is not well-suited for light nuclei due to large fluctuations in the $Z<20$ region. It is worth noting that our estimations tend to overestimate true values of $R_\textrm{x}$, and the discrepancy increases with $Z$, yet the deviations are typically less than 1\% for most nuclei of interest. We used this estimate when computing the uncertainties.

\section{Results and Discussion}
\label{sec:Results}

In the following, we present extensive comparisons between our results and the experimental values for various relative electron capture ratios, such as $\lambda_K/\lambda$, $\lambda_L/\lambda_K$, etc. If not specified otherwise, the experimental values are taken from the tables of recommended data (RD) from Refs. \cite{TabRad_v0, TabRad_v1, TabRad_v2, TabRad_v3, TabRad_v4, TabRad_v5, TabRad_v6, TabRad_v7, TabRad_v8}. When experimental values available, we denote all upper shells using the plus notation, e.g. "L+" denotes all the upper shells starting with the L shell. We used the notation "UF" for unique forbidden transitions. The $Q$-values for each electron capture process is taken from the \textsc{AME2020} database \cite{AME2020} and the excitation energy of the level populated in the final nucleus, denoted in this paper as $R_{\gamma}$, is taken from \cite{livechart}. When a comparison is done with the previous theoretical models, the energy $Q-R_\gamma$ is taken as in the reference(s) presenting the models. We mention that if not specified otherwise, the more precise energetics (see Eq.~\ref{eq:NewNeutrinoEnergy}) was implied in the calculation of the electron capture fractions. 

The assessment of uncertainties for the capture ratios is discussed in the Appendix. Here, we mention that in the BS and KLI  models the uncertainties are solely attributed to $Q$ and $R_{\gamma}$, as stated in \cite{MougeotARI2018}. However, the uncertainty of the atomic structure calculation, which plays a crucial role in determining the emitted neutrino energy and, consequently, the decay rate, is not considered. To address this limitation, our model introduces a solution by incorporating an associated uncertainty to the atomic relaxation energies. Thus, we provide a more comprehensive assessment of the overall uncertainty in electron capture observables. Moreover, the ad-hoc assessment of uncertainties \cite{MougeotARI2018} does not guarantee 68\% coverage. Our method, using pseudo-experiments, meets the coverage criterion.

\begin{table*}
	\begin{center}
		\begin{tabular}{cccccccccc}
			\hline \hline Isotope & $Q-R_\gamma$(keV)\cite{MougetasrXiv2021} & Type & Quantity & BS\cite{MougetasrXiv2021} & KLI\cite{MougetasrXiv2021}  & KLI\cite{MougetasrXiv2021}  & This work \footnotemark[1] & This work \footnotemark[2] & RD\cite{TabRad_v0, TabRad_v1, TabRad_v2, TabRad_v3, TabRad_v4, TabRad_v5, TabRad_v6, TabRad_v7, TabRad_v8} \\
			& & & & & no vacancy & frozen orbitals & & & \\
			\hline${ }^{7} \mathrm{Be}$ & $861.89(7)$ & Allowed & $\lambda_{L} / \lambda_{K}$ & $0.105(8)$ & $0.1606(41)$ & $0.0509(20)$ & $0.11054(3)$ &$0.11053(3)$ & $0.101(13)$ \\
			\hline${ }^{41} \mathrm{Ca}$ & $421.64(14)$ & 1st UF & $\lambda_{L} / \lambda_{K}$ &$0.09800(40)$ & $0.10415(16)$ & $0.09078(16)$ & $0.1050(2)$ & $0.1046(2)$ & $0.102(10)$ \\
			\hline${ }^{54} \mathrm{Mn}$ & $542.2(10)$ & Allowed & $\lambda_{L} / \lambda_{K}$ & $0.11219(31)$ & $0.10785(8)$ & $0.09590(19)$ & $0.1078(6)$ & $0.1076(6)$ & $0.1066(16)$ \\
			& & & $\lambda_{K}/\lambda$ & $0.88419(34)$ & $0.88623(10)$ & $0.90005(21)$ & $0.8869(5)$ & $0.8870(5)$ & $0.8896(17)$ \\
			\hline${ }^{55} \mathrm{Fe}$ & $231.21(18)$ & Allowed & $\lambda_{L} / \lambda_{K}$ & $0.11629(31)$ & $0.11236(8)$ & $0.10073(20)$ & $0.1125(3)$ & $0.1121(3)$ & $0.1110(15)$ \\
			& & & $\lambda_{M} / \lambda_{K}$ & $0.01824(12)$ & $0.019390(32)$ & $0.014824(45)$ & $0.01918(4)$ & $0.01909(5)$ & $0.01786(29)$ \footnotemark[3] \\
			& & & $\lambda_{M} / \lambda_{L}$ & $0.1568(11)$ & $0.17257(31)$ & $0.14716(49)$ & $0.1705(4)$ & $0.1704(4)$ &  $0.1556(26)$ \footnotemark[3]\\
			\hline${ }^{109} \mathrm{Cd}$ & $127.1(18)$ & Allowed & $\lambda_{K}/\lambda$ & $0.8148(14)$ & $0.8097(11)$ & $0.8164(12)$ & $0.807(7)$ & $0.810(7)$ & $0.812(3)$ \\
			& & & $\lambda_{L+} / \lambda_{K}$ & $0.2274(12)$ & $0.2350(11)$ & $0.2250(12)$ & $0.2390(101)$ & $0.2344(101)$ & $0.2315(8)$ \\
			\hline${ }^{125} \mathrm{I}$ & $150.28(6)$ & Allowed & $\lambda_{K}/\lambda$ & $0.79927(41)$ & $0.79798(7)$ & $0.80376(23)$ & $0.7952(2)$ & $0.7983(18)$ & $0.8011(17)$ \\
			\hline${ }^{138} \mathrm{La}$ & $312.6(3)$ & 2nd UF & $\lambda_{L} / \lambda_{K}$ & $0.3913(25)$ & $0.4077(15)$ & $0.4242(49)$ & $0.420(3)$ & $0.409(7)$ & $0.432(6)$ \\
			& & & $\lambda_{M} / \lambda_{K}$ & $0.0965(9)$ & $0.09908(41)$ & $0.1002(11)$ & $0.1025(8)$ & $0.100(2)$ & $0.102 (3)$ \footnotemark[3] \\
			& & & $\lambda_{M} / \lambda_{L}$ & $0.2465(20)$ & $0.2430(22)$ & $0.2362(24)$ & $0.244(1)$ & $0.244(1)$ & $0.261(9)$ \footnotemark[3] \\
			\hline \hline
		\end{tabular}
	\end{center}
	\footnotetext[1]{the neutrino energy is determined using the approximate energetics from Eq.~\ref{eq:OldNeutrinoEnergy}}
	\footnotetext[2]{the neutrino energy is determined using the more precise energetics from Eq.~\ref{eq:NewNeutrinoEnergy}}
	\footnotetext[3]{from Ref.~\cite{MougetasrXiv2021}}
	\caption{\label{tab:Models} Comparison between several theoretical models results and measured electron capture decay ratios for isotopes studied in \cite{MougetasrXiv2021}.}
\end{table*}

In this study, we investigate the electron capture shell ratios and relative ratios of multiple nuclei, and compare them with other theoretical models (BS and KLI) as well as experimental data. Table \ref{tab:Models} displays the results for seven different transitions, including five allowed, one first unique forbidden, and one second unique forbidden transition, which span a wide range of mass numbers. The KLI model is evaluated using two separate cases: the "no vacancy" approximation and the "frozen orbitals" approximation. The former assumes that the final nucleus is in the ground state configuration, while the latter assumes that the atomic configuration of the final atom is the same as the initial one, minus the captured electron. In order to have a consistent comparison between models, we use the approximate energetics presented in Eq.~\ref{eq:OldNeutrinoEnergy}. All models predict $\lambda_{K}/\lambda$ ratios within 2\% of experimental values. However, the deviation increases to 12\% when comparing $\lambda_{L}/\lambda_{K}$ ratios, which suggests large differences between models in the computation of $\lambda_{L}$. Other predicted ratios are within 10\% of experimental values. All quoted deviations are quite large when compared with the experimental uncertainties. Nonetheless, our model and BS model provide the most accurate predictions. Next, we present the results with refined energetics. The main effect of this change is a smaller deviation from experimental values for most fractions, especially for the low $Q-R_{\gamma}$ transitions of $^{109}$Cd and $^{125}$I. We will come back to the comparison between energetics with a graphical representation.

\begin{table*}
	\begin{center}
		\begin{tabular}{ccccccc}
			\hline \hline Isotope & $Q$(keV)\cite{AME2020} & $R_\gamma$(keV)\cite{livechart} & Type & Quantity & This work & RD\cite{TabRad_v0, TabRad_v1, TabRad_v2, TabRad_v3, TabRad_v4, TabRad_v5, TabRad_v6, TabRad_v7, TabRad_v8} \\
			
			\hline${ }^{7} \mathrm{Be}$ & $861.89(7)$ & $0$ & Allowed & $\lambda_K/\lambda$ & $0.90047(2)$ & $0.908(12)$ \\
			& & & & $\lambda_L/\lambda$ & $0.09952(2)$ & $0.092(12)$ \\
			& $861.89(7)$ & $477.612(3)$ & Allowed & $\lambda_K/\lambda$ & $0.90046(5)$ & $0.908(12)$ \\
			& & & & $\lambda_L/\lambda$ & $0.09954(5)$ & $0.092(12)$ \\
			
			\hline${ }^{22} \mathrm{Na}$ & $2843.32(13)$ & $1274.537(7)$ & Allowed & $\lambda_K/\lambda$ & $0.91925(2)$ & $0.923(4)$ \\
			& & & & $\lambda_L/\lambda$ & $0.07877(2)$ & $0.077(4)$ \\
			
			\hline${ }^{37} \mathrm{Ar}$ & $813.87(20)$ & $0$ & Allowed & $\lambda_K/\lambda$ & $0.89892(6)$ & $0.9021(24)$ \\
			& & & & $\lambda_L/\lambda$ & $0.08853(6)$ & $0.0872(20)$ \\
			& & & & $\lambda_M/\lambda$ & $0.012551(9)$ & $0.0106(7)$ \\
			
			\hline${ }^{51} \mathrm{Cr}$ &$752.39(15)$ & $0$ & Allowed & $\lambda_K/\lambda$ & $0.89004(6)$ & $0.8919(17)$ \\
			& & & & $\lambda_L/\lambda$ & $0.09427(5)$ & $0.0927(14)$ \\
			& & & & $\lambda_M/\lambda$ & $0.015688(9)$ & $0.0154(6)$ \\
			& $752.39(15)$ & $320.0835(4)$ & Allowed & $\lambda_K/\lambda$ & $0.8891(1)$ & $0.8910(17)$ \\
			& & & & $\lambda_L/\lambda$ & $0.09509(9)$ & $0.0935(14)$ \\
			& & & & $\lambda_M/\lambda$ & $0.01584(2)$ & $0.0156(6)$ \\ 
			
			\hline${ }^{64} \mathrm{Cu}$ & $1674.62(21)$ & $1345.777(23)$ & Allowed & $\lambda_K/\lambda$ & $0.8815(2)$ & $0.884(3)$ \\
			& & & & $\lambda_L/\lambda$ & $0.1008(2)$ & $0.099(2)$ \\
			& & & & $\lambda_M/\lambda$ & $0.01714(3)$ & $0.0162(5)$ \\
			
			\hline${ }^{73} \mathrm{Se}$ & $2725(7)$ & $427.902(21)$ & Allowed & $\lambda_K/\lambda$ & $0.8792(9)$ & $0.8810(15)$ \\
			& & & & $\lambda_L/\lambda$ & $0.1006(8)$ & $0.1001(12)$ \\
			& & & & $\lambda_M/\lambda$ & $0.0180(1)$ & $0.0172(4)$ \\
			
			\hline${ }^{88} \mathrm{Y}$ & $3622.6(15)$ & $3584.784(19)$ & Allowed & $\lambda_K/\lambda$ & $0.716(34)$ & $0.721(12)$ \\
			& & & & $\lambda_L/\lambda$ & $0.227(28)$ & $0.225(10)$ \\
			& & & & $\lambda_M/\lambda$ & $0.0475(63)$ & $0.0542(25)$ \\
			
			\hline${ }^{108} \mathrm{Ag}$ & $1917.4(2.6)$ & $1052.78(5)$ & Allowed & $\lambda_K/\lambda$ & $0.8595(10)$ & $0.8611(14)$ \\
			& & & & $\lambda_L/\lambda$ & $0.1122(9)$ & $0.1118(11)$ \\
			& & & & $\lambda_M/\lambda$ & $0.0232(2)$ & $0.0227(5)$ \\
			
			\hline${ }^{124} \mathrm{I}$ & $3159.6(19)$ & $2335.03(1)$ & 2nd UF & $\lambda_K/\lambda$ & $0.82040(273)$ & $0.82099(43)$ \\
			& & & & $\lambda_L/\lambda$ & $0.14024(233)$ & $0.13959(19)$ \\
			& & & & $\lambda_M/\lambda$ & $0.03099(54)$ & $0.03135(15)$ \\
			& $3159.6(19)$ &$2483.362(13)$ & 1st UF & $\lambda_K/\lambda$ & $0.83148(215)$ & $0.83184(41)$ \\
			& & & & $\lambda_L/\lambda$ & $0.13175(183)$ & $0.13127(18)$ \\
			& & & & $\lambda_M/\lambda$ & $0.02891(43)$ & $0.02928(14)$ \\
			
			\hline${ }^{142} \mathrm{Pr}$ & $746.5(25)$ & $0$ & 1st UF & $\lambda_K/\lambda$ & $0.8373(13)$ & $0.8398(15)$ \\
			& & & & $\lambda_L/\lambda$ & $0.1258(11)$ & $0.1255(11)$ \\
			& & & & $\lambda_M/\lambda$ & $0.0283(3)$ & $0.0280(5)$ \\
			
			\hline${ }^{152} \mathrm{Eu}$ & $1874.5(7)$ & $1529.802(3)$ & Allowed & $\lambda_K/\lambda$ & $0.8082(13)$ & $0.8109(17)$ \\
			& & & & $\lambda_L/\lambda$ & $0.1467(10)$ & $0.1465(12)$ \\
			& & & & $\lambda_M/\lambda$ & $0.0345(3)$ & $0.0341(7)$ \\
			
			\hline${ }^{169} \mathrm{Yb}$ & $899.1(8)$ & $316.14633(11)$ & Allowed & $\lambda_K/\lambda$ & $0.8074(9)$ & $0.8093(17)$ \\
			& & & & $\lambda_L/\lambda$ & $0.1465(7)$ & $0.1457(12)$ \\
			& & & & $\lambda_M/\lambda$ & $0.0353(2)$ & $0.0349(7)$ \\
			
			\hline${ }^{195} \mathrm{Au}$ & $226.8(10)$ & $98.880(2)$ & 1st UF & $\lambda_K/\lambda$ & $0.445(19)$ & $0.452(6)$ \\
			& & & & $\lambda_L/\lambda$ & $0.400(14)$ & $0.398(4)$ \\
			& & & & $\lambda_{M+}/\lambda$ & $0.1552(53)$ & $0.1499(18)$ \\
			
			\hline${ }^{204} \mathrm{Tl}$ & $344.1(12)$ & $0$ & 1st UF & $\lambda_K/\lambda$ & $0.5845(78)$ & $0.5843(14)$ \\
			& & & & $\lambda_L/\lambda$ & $0.3016(59)$ & $0.3024(10)$ \\
			& & & & $\lambda_{M+}/\lambda$ & $0.1138(22)$ & $0.1133(5)$ \\	
			
			\hline \hline
		\end{tabular}
	\end{center}
	\caption{Systematic comparison between measured electron capture decay ratios and our predictions for selected isotopes. The precise energetics was used in computing values in the sixth column.}
	\label{tab:P-all}
\end{table*}

We now turn evaluate the validity of our model across a wide range of atomic numbers and various types of transitions. In Table~\ref{tab:P-all}, we computed the relative electron capture ratios for a set of nuclei spanning from $^{7}$Be to $^{204}$Tl. In most cases the predictions of $\lambda_{K}/\lambda$ ratios agree with experimental values within one standard deviation. The largest discrepancy in this quantity is still around two standard deviations, for example the second unique forbidden transition of $^{124}$I and the allowed transition of $^{152}$Eu. For the majority of cases, our results were consistent with the recommended data within two standard deviations. The only instances where a simple $\chi^2$ test with one degree of freedom fails at 90\% confidence level the $\lambda_{M}/\lambda$ ratios of $^{37} $Ar, $^{64}$Cu and $^{73}$Se. In a few instances where high-precision RD were available ($^{124}$I, $^{152}$Eu, $^{204}$Tl), the deviation of the central values is within 3\%. However, the theoretical uncertainty is quite large (compared to the experimental one) in these cases and compatibility between theory and experiment is not affected. Overall, our calculations were in excellent agreement with the experimental data. 

\begin{table*}
	\begin{center}
		\begin{tabular}{ccccccccc}
			\hline \hline Isotope & $Q$(keV)\cite{AME2020} & $R_\gamma$(keV)\cite{livechart} & & $\lambda_{K}/\lambda$ & $\lambda_{L}/\lambda$ & $\lambda_{M}/\lambda\times10^{2}$ & $\lambda_{N}/\lambda \times10^{3}$ & $\lambda_{O}/\lambda \times10^{3}$\\
			
			\hline
			${ }^{67} \mathrm{Ga}$ & $1001.2(11)$ & $0$ & This work & $0.8818(3)$ & $0.0995(3)$ & $1.715(5)$ & $1.504(5)$ & $0$\\
			& & & RD & $0.8836(15)$ &$0.0989(12)$ & $1.640(40)$ & - &\\
			
			\hline
			${ }^{67} \mathrm{Ga}$ & $1001.2(11)$ & $93.312(5)$ & This work & $0.8816(3)$ & $0.0997(3)$	& $1.718(6)$ & $1.507(5)$ & $0$\\
			& & & RD & $0.8834(15)$ &$0.0991(12)$ & $1.640(40)$& - &\\
			
			\hline
			${ }^{67} \mathrm{Ga}$ & $1001.2(11)$ & $184.579(6)$ & This work & $0.8814(4)$ & $0.1000(3)$ & $1.722(6)$ & $1.510(6)$ & $0$\\
			& & & RD & $0.8832(15)$ &$0.0993(12)$ & $1.640(40)$ & - &\\
			
			\hline			
			${ }^{67} \mathrm{Ga}$ & $1001.2(11)$ & $393.531(7)$ & This work & $0.8806(5)$ & $0.1005(4)$ & $1.734(8)$ & $1.521(8)$ & $0$\\
			& & & RD & $0.8824(15)$ &$0.0999(12)$ & $1.650(40)$ & - &\\
			
			\hline
			${ }^{67} \mathrm{Ga}$ & $1001.2(11)$ & $887.701(8)$ & This work & $0.8661(32)$ & $0.1125(29)$ &$1.970(53)$ & $1.730(47)$ & $0$\\
			& & & RD & $0.8680(17)$ &$0.1119(14)$ & $1.88(5)$ & - &\\
			
			\hline
			${ }^{111} \mathrm{In}$ & $860(3)$ & $416.72(3)$ & This work & $0.8494(24)$ & $0.11922(206)$ & $2.520(46)$ & $5.532(101)$ & $0.609(11)$\\
			& & & RD & $0.8518(2)$ &$0.11835(13)$ & - & - & - \\
			
			\hline
			${ }^{123} \mathrm{I}$ & $1228(3)$ & $158.994(22)$ & This work & $0.8513(10)$ & $0.1166(8)$ & $2.516(2)$ & $5.913(45)$ & $0.942(7)$\\
			& & & RD & $0.8533(14)$ & $0.1163(10)$ & $2.48(5)$& - & - \\
			
			\hline
			${ }^{123} \mathrm{I}$ & $1228(3)$ & $440.00(4)$ & This work & $0.8489(13)$ & $0.1185(11)$ & $2.563(26)$ & $6.027(61)$ & $0.960(10)$\\
			& & & RD & $0.8510(14)$ &$0.1181(10)$ & $2.53(5)$& - & - \\
			
			\hline
			${ }^{123} \mathrm{I}$ & $1228(3)$ & $489.78(5)$ & This work & $0.8482(14)$ & $0.1190(12)$ & $2.580(28)$ & $6.057(66)$ & $0.965(11)$\\
			& & & RD & $0.8503(14)$ & $0.1186(10)$ & $2.54(5)$ & - & - \\
			
			\hline
			${ }^{123} \mathrm{I}$ & $1228(3)$ & $505.35(4)$ & This work & $0.8480(14)$ & $0.1191(13)$ & $2.580(28)$ & $6.067(68)$ & $0.966(11)$\\
			& & & RD & $0.8501(14)$ & $0.1187(10)$ & $2.54(5)$& - & - \\
			
			\hline
			${ }^{123} \mathrm{I}$ & $1228(3)$ & $687.97(3)$ & This work & $0.8444(20)$ & $0.1219(17)$ & $2.649(40)$ & $6.233(94)$ & $0.993(15)$\\
			& & & RD & $0.8464(14)$ &$0.1216(10)$ & $2.62(5)$& - & -\\
			
			\hline
			${ }^{123} \mathrm{I}$ & $1228(3)$ & $783.62(3)$ & This work & $0.8412(25)$ & $0.1243(21)$	& $2.709(49)$ & $6.380(117)$ & $1.017(19)$\\
			& & & RD & $0.8436(14)$ & $0.1237(10)$ & $2.67(5)$& - & - \\
			
			\hline
			${ }^{123} \mathrm{I}$ & $1228(3)$ & $894.77(6)$ & This work & $0.8350(35)$ & $0.1290(30)$ & $2.829(69)$ & $6.671(163)$ & $1.064(26)$ \\
			& & & RD & $0.8377(15)$ & $0.1283(11)$ & $2.78(5)$ & - & - \\
			
			\hline
			${ }^{123} \mathrm{I}$ & $1228(3)$ & $1036.62(5)$ & This work & $0.8140(72)$ & $0.1448(60)$	& $3.234(140)$ & $7.652(333)$ & $1.221(54)$\\
			& & & RD & $0.8182(18)$ &$0.1427(13)$ & $3.16(6)$& - & - \\
			
			\hline
			${ }^{123} \mathrm{I}$ & $1228(3)$ & $1068.23(6)$ & This work & $0.8028(93)$ & $0.1532(76)$	& $3.450(180)$ & $8.179(433)$ & $1.305(70)$\\
			& & & RD & $0.8082(21)$ &$0.1503(15)$ & $3.36(7)$& - & - \\
			
			\hline
			${ }^{125} \mathrm{I}$ & $185.77(6)$ & $35.4925(5)$ & This work & $0.7983(18)$ & $0.1566(14)$ &$3.539(31)$& $8.395(74)$& $1.340(12)$ \\
			& & & RD & $0.8011(17)$ & $0.1561(13)$ & $3.49(7)$& - & - \\
			
			\hline
			${ }^{125} \mathrm{Xe}$ & $1636.7(4)$ & $188.416(4)$ & This work & $0.8516(1)$ & $0.1162(1)$ & $2.519(3)$ & $6.011(6)$ & $1.010(1)$ \\
			& & & RD & - & - & - & - & - \\
			
			\hline
			${ }^{125} \mathrm{Xe}$ & $1636.7(4)$ & $243.382(4)$ & This work & $0.8514(2)$ & $0.1163(1)$ & $2.523(3)$ & $6.021(7)$ & $1.011(1)$ \\
			& & & RD & - & - & - & - & - \\
			
			\hline
			${ }^{125} \mathrm{Xe}$ & $1636.7(4)$ & $372.066(13)$ & This work & $0.8508(2)$ & $0.1168(1)$ & $2.533(3)$ & $6.047(7)$ & $1.016(1)$ \\
			& & & RD & - & - & - & - & - \\
			
			\hline
			${ }^{125} \mathrm{Xe}$ & $1636.7(4)$ & $453.792(9)$ & This work & $0.8504(2)$ & $0.1171(1)$ & $2.541(3)$ & $6.066(8)$ & $1.019(1)$ \\
			& & & RD & - & - & - & - & - \\		
			
			\hline
			${ }^{125} \mathrm{Xe}$ & $1636.7(4)$ & $1007.450(16)$ & This work & $0.8446(4)$ & $0.1215(3)$ & $2.653(6)$ & $6.342(16)$ & $1.066(3)$ \\
			& & & RD & - & - & - & - & - \\
			
			\hline
			${ }^{125} \mathrm{Xe}$ & $1636.7(4)$ & $1089.904(12)$ & This work & $0.8426(4)$ & $0.1230(3)$ & $2.691(8)$ & $6.434(18)$ & $1.081(3)$ \\
			& & & RD & - & - & - & - & - \\
			
			\hline
			${ }^{125} \mathrm{Xe}$ & $1636.7(4)$ & $1180.872(13)$ & This work & $0.8396(5)$ & $0.1253(4)$ & $2.750(9)$ & $6.579(22)$ & $1.106(4)$ \\
			& & & RD & - & - & - & - & - \\
			
			\hline
			${ }^{125} \mathrm{Xe}$ & $1636.7(4)$ & $1442.79(5)$ & This work & $0.8109(15)$ & $0.1468(12)$ & $3.304(28)$ & $7.948(68)$ & $1.337(11)$ \\
			& & & RD & - & - & - & - & - \\
			
			\hline \hline
		\end{tabular}
	\end{center}
	\caption{\label{tab:P-med-XE1T} Systematic comparison between our theoretical predictions and experimental values of the capture ratios for selected nuclei of interest in nuclear medicine and exotic physics searches with liquid Xenon detectors.}
\end{table*}

We now focus on the allowed electron capture (EC) transitions of five nuclei ($^{67}\mathrm{Ga}$, $^{111}\mathrm{In}$, $^{123}\mathrm{I}$, $^{125}\mathrm{I}$, $^{125}\mathrm{Xe}$), which are of practical interest in either nuclear medicine or exotic physics searches in liquid xenon experiments. It is worth noting that we could not find any recommended data (RD) for $^{125}$Xe in the references mentioned above. Therefore, we present only the theoretical predictions for all excited states populated in the final nucleus, considering their significance in the construction of background models in liquid Xenon experiments. The predicted relative capture ratios are shown and compared with experimental values, where available, in Table~\ref{tab:P-med-XE1T}. As before, our model predicts $\lambda_{K}/\lambda$ ratios in excellent agreement (below two standard deviations) with experimental measurements. Remarkably, even better agreement is obtained for $\lambda_{L}/\lambda$ ratios (below one standard deviation in most cases). This feature is a consequence of the improved energetics. Deviations between experimental and theoretical predictions for $\lambda_{M}/\lambda$ ratios are slightly larger (still within three sigma) only for the first four transitions of $^{67}$Ga. Nonetheless, we expect our model to provide reliable predictions for all interesting nuclei. For completeness, we also include the $\lambda_{N}/\lambda$ and $\lambda_{O}/\lambda$ ratios, although no experimental values are available.

\begin{figure*}
	\centering
	\includegraphics[width=0.8\linewidth]{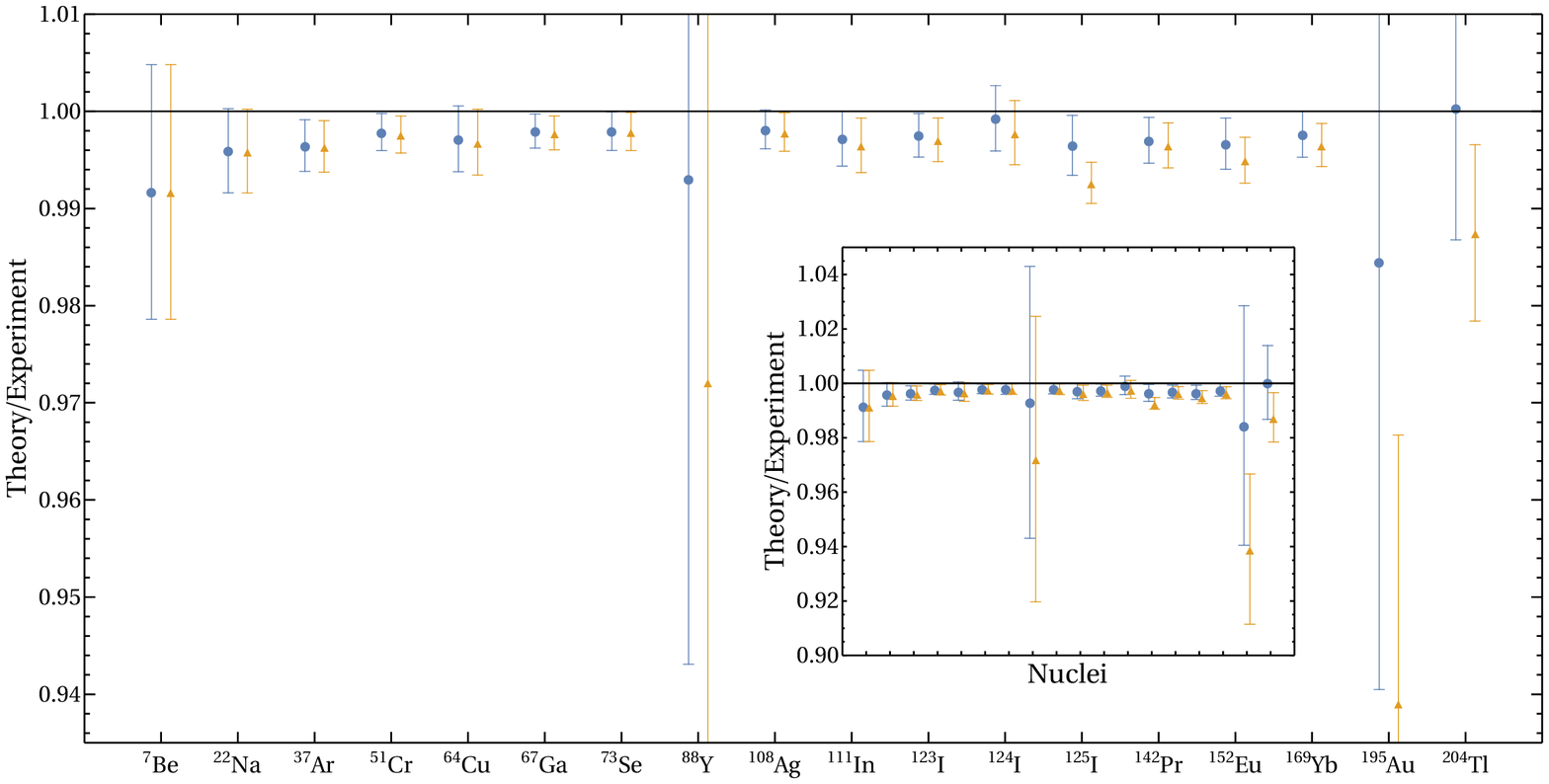}
	\caption{\label{fig:RapPxK} The ratio between the theoretical and experimental values for relative capture probability from the $K$ shell considering both the approximate (orange triangle) and the refined (blue circle) method of computing the neutrino energy. The inset is the same figure but with extended $y$-axis. Error bars are computed with theoretical and experimental uncertainties summed in quadrature.}
\end{figure*}

\begin{figure*}
	\centering
	\includegraphics[width=0.8\linewidth]{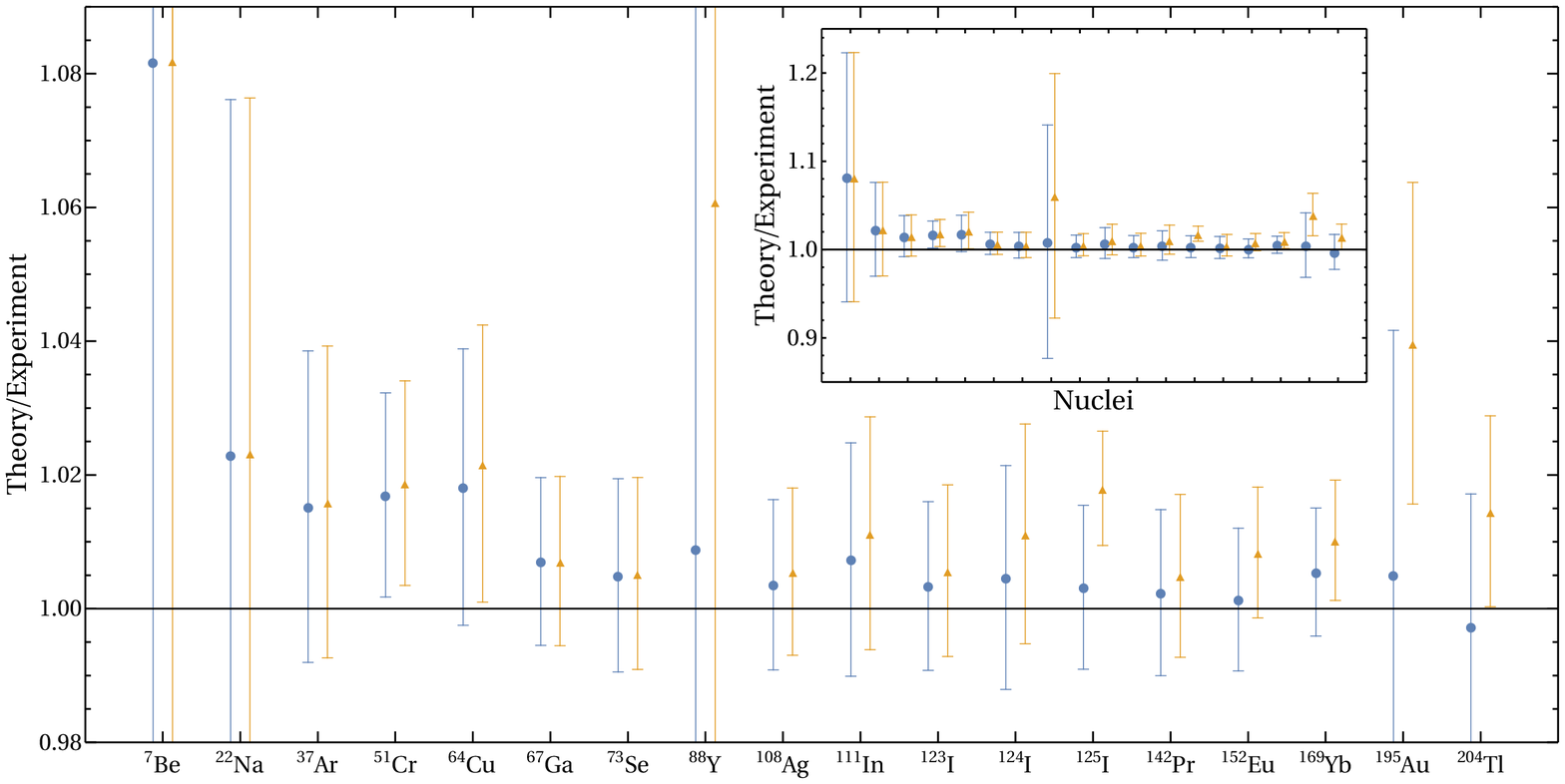}
	\caption{\label{fig:RapPxL}	Same as Fig.~\ref{fig:RapPxK} but for the $L$ shell}
\end{figure*}

Finally, we discuss the effect of the energetics refinement. To this purpose we plot in Fig.~\ref{fig:RapPxK} and~\ref{fig:RapPxL} the $\lambda_{K}/\lambda$ and $\lambda_{L}/\lambda$ ratios respectively, for all nuclei presented in Tables~\ref{tab:P-all} and~\ref{tab:P-med-XE1T}. For transitions with multiple possible nuclear final states, we chose the following: $^7$Be$ \to $ ground state, $^{51}$Cr$ \to 320.0835$~keV, $^{67}$Ga$ \to$ 184.579~keV, $^{123}$I$ \to$ 440~keV, $^{124}$I$ \to$ 2335.03~keV. For each transition, we employ both the approximate and the refined energetics (see Eq.~\ref{eq:OldNeutrinoEnergy} and Eq.~\ref{eq:NewNeutrinoEnergy} respectively). Our results show that theoretical predictions based on refined energetics exhibit better agreement with experimental data, as evidenced by an increase in $\lambda_{K}/\lambda$ ratios and a decrease in $\lambda_{L}/\lambda$ ratios, for all nuclei studied. The improvement is more significant for transitions with low values of $Q - R_{\gamma}$, as expected. In particular, we highlight the transitions of $^{125}$I, $^{152}$Eu, $^{195}$Au and $^{204}$Tl, with $Q-R_{\gamma}=150.3$ keV, 344.7 keV, 127.92 keV and 344.1 keV, respectively. Theoretical estimations based on approximate energetics deviate from experimental values by more than one standard deviation. However, employing refined energetics shifts the theoretical predictions to almost perfect agreement, even at the level of central values.

\section{Conclusions}
\label{sec:Conslusions}

We presented a thorough re-examination of the EC formalism and calculations of capture fractions, including several critical atomic effects. We employed the Dirac-Hartree-Fock-Slater self-consistent framework for the bound electron wave functions. The reliability of this method for EC calculations is underlined by systematic comparisons of binding energies and Coulomb amplitudes with previous theoretical models and experimental data. 

The novel aspect of this paper involves a refined evaluation of the energy of the emitted neutrino. Previous models determine the emitted neutrino energy by neglecting the total change in electron binding energy between initial and final atoms and the rearrangement energy of the captured electron. We propose a more rigorous method, in which both the above quantities are computed using atomic structure calculations. Although CPU-intensive, this approach yields improved agreement with experimental values for electron capture fractions, particularly in low energy transitions.

In the comparison between theoretical predictions we observe deviations from experimental values below 2\% for the $\lambda_{K}/\lambda$ ratios for all models. These deviations increase to 12\% for captures from higher shells. Overall, the BS model and our model provide the most accurate values. Furthermore, we tested the validity of our model by comparing theoretical predictions to experimental values across wide ranges of atomic numbers and transition energies. We found excellent agreement (below two standard deviations) between theory and experiment in most cases. The only exceptions are the $\lambda_{M}/\lambda$ ratios of $^{37} $Ar, $^{64}$Cu and $^{73}$Se. We also investigated in detail five nuclei ($^{67}\mathrm{Ga}$, $^{111}\mathrm{In}$, $^{123}\mathrm{I}$, $^{125}\mathrm{I}$, $^{125}\mathrm{Xe}$) of interest in either nuclear medicine or exotic physics searches with liquid Xenon detectors. We obtained less than one standard deviation between predictions and experimental values of the relative capture fractions in almost all cases, except $\lambda_{M}/\lambda$ for the first four excited final states of the $^{67}$Ga decays.

Finally, we note that using refined energetics leads to a better agreement between experimental and theoretical EC ratios, particularly for low-energy transitions. We believe that this characteristic could have an impact on the determination of the neutrino mass scale from electron capture processes. In conclusion, our results have significant implications for future studies in the field of EC, as well as for related applications in nuclear physics and astrophysics.

\section*{ACKNOWLEDGMENTS}

This work was supported by grants of Romanian Ministry of Research, Innovation and Digitalization through the project CNCS – UEFISCDI
No. 99/2021 within PN-III-P4-ID-PCE-2020-2374, the
project CNCS – UEFISCDI No. TE12/2021 within PN-III-
P1-1.1-TE-2021-034.

The figures for this article have been created using the SciDraw scientific figure preparation system \cite{SciDraw}.

\appendix

\section{Uncertainties estimation}
In this study, we assess the uncertainties associated with the capture ratios presented in tables~\ref{tab:Models},~\ref{tab:P-all}, and~\ref{tab:P-med-XE1T} using a pseudo-experiment technique. The approach involves the sampling of parameters that can potentially fluctuate from probability distribution functions (pdfs). The capture rates are then computed for each set of sampled parameters, and the corresponding ratios ($r$) are determined, resulting in pdfs for the ratios ($f(r)$). The mean values ($\hat{r}$) of these pdfs are reported as the central values in the tables. To obtain the two-sided bounds ($r_\textrm{low}$ and $r_\textrm{up}$) for each ratio, we solve the following equations
\begin{align}
	\begin{aligned}
		\int_{0}^{r_{\textrm{low}}}drf(r) &= 0.16,\\
		\int_{r_\textrm{up}}^{1}drf(r) &=0.16.
	\end{aligned}
\end{align}
Since the pdfs $f(r)$ are not symmetric, we avoid under-coverage by quoting as uncertainty the number
\begin{equation}
	\max(r-r_\textrm{low},r_\textrm{up}-r).
\end{equation}
This procedure has the advantage of correctly accounting for correlations between the numerator and denominator of each ratio. 

The primary factor contributing to the uncertainty in the capture rate ratios is the neutrino energy. To account for this, we only sample $q_\textrm{x}$ in Eq.~\ref{eq:NewNeutrinoEnergy} as follows. We assume that the $Q$-value and nuclear relaxation energy $R_\gamma$ are normally distributed with experimental values as means and standard deviations. The theoretical uncertainty in the atomic relaxation energy $R_\textrm{x}$ is the only remaining source of uncertainty. In Fig.~\ref{fig:edge_energies_dhfs} we show that the DHFS self-consistent method agrees with experimental values within 3\% for all $R_{\textrm{x}}$. Therefore, we use a uniform distribution to model $R_\textrm{x}$ in our pseudo-experiments, with a range of $[0.97 \hat{R}_\textrm{x}, 1.03\hat{R}_\textrm{x}]$, where $\hat{R}_{\textrm{x}}$ is obtained from Eq.~\ref{eq:EnergyDepositWithBindingEnergy}.

\begin{figure}
	\includegraphics[width=\columnwidth]{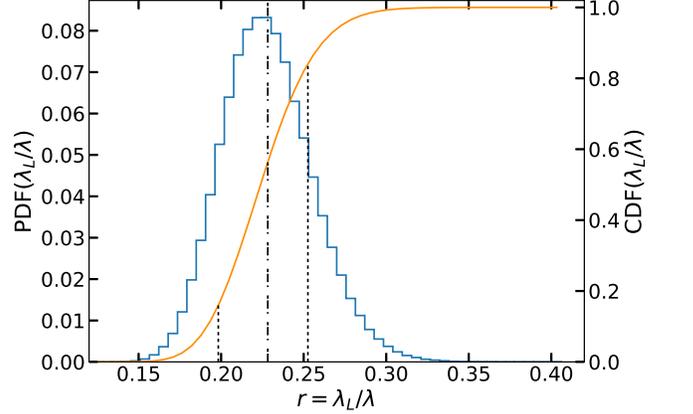}
	\caption{\label{fig:r_pdf} Result of $10^6$ pseudo-experiments used for the evaluation of the uncertainty of $r\lambda_L/\lambda$ for the allowed $e^- + ^{88}\textrm{Y} \to ^{88}\textrm{Sr}^{*} + \nu_e$ transition. The blue histogram is the pdf of $r$. The orange curve is the cumulative distribution function of $r$. The vertical dot-dashed line passes through the mean value $\hat{r}$. The two dotted lines correspond to $r_\textrm{low}$ and $r_\textrm{up}$.}
\end{figure}

For each nucleus, we use $10^6$ pseudo-experiments. For illustration purposes, the results for the ratio $r=\lambda_{L}/\lambda$ for the allowed transition $e^- + ^{88}\textrm{Y} \to ^{88}\textrm{Sr}^{*} + \nu_e$ are shown in Fig.~\ref{fig:r_pdf}.

\bibliographystyle{apsrev4-2}
\bibliography{apssamp}

\end{document}